\documentclass[traditabstract]{aa}
\usepackage[dutch, english]{babel}
\usepackage{amssymb}
\usepackage{amsmath}
\usepackage{mathrsfs}
\usepackage{latexsym}
\usepackage{longtable}
\usepackage{multirow}
\usepackage{epsf}
\usepackage{color}
\usepackage{graphicx}
\usepackage{float}
\usepackage{enumerate}
\usepackage{txfonts}
\usepackage{natbib}

\newcommand{\kms}{km~s$^{-1}$ }
\newcommand{\ergscm}{erg~s$^{-1}$~cm$^{-2}$ }
\newcommand{\one}{~{\sc i}}
\newcommand{\two}{~{\sc ii}}
\newcommand{\three}{~{\sc iii}}

\title{The outflow history of two Herbig-Haro jets in \object{RCW~36}:\\ \object{HH~1042} and \object{HH~1043}\thanks{Based on observations performed with X-shooter (program P87.C-0442) mounted on the ESO {\it Very Large Telescope} on Cerro Paranal,  Chile}}

\authorrunning{L.E. Ellerbroek et al.}
\titlerunning{The history of HH~1042 and HH~1043}

\author{L.~E.~Ellerbroek \inst{1}
\and
L.~Podio\inst{2,3}
\and
L.~Kaper\inst{1}
\and
H.~Sana\inst{1}
\and
D.~Huppenkothen\inst{1}
\and
A.~de~Koter\inst{1,4,5}
\and
L.~Monaco\inst{6}
}

\institute{Astronomical Institute ``Anton Pannekoek'', University of Amsterdam, Science Park 904, 1098 XH Amsterdam, The Netherlands\\
\email{l.e.ellerbroek@uva.nl}
\and
Institut de Plan\'{e}tologie et d'Astrophysique de Grenoble, 414, Rue de la Piscine, 38400 St-Martin d'H\`{e}res, France
\and
Kapteyn Institute, Landleven 12, 9747 AD Groningen, The Netherlands
\and
Astronomical Institute, Utrecht University, Princetonplein 5, 3584 CC Utrecht, The Netherlands
\and
Instituut voor Sterrenkunde, KU Leuven, Celestijnenlaan 200B, 3001 Leuven, Belgium
\and 
European Southern Observatory, Alonso de Cordova 3107, Casilla 19001, Santiago, Chile
}

\date{Received; accepted}

\abstract{Jets around low- and intermediate-mass young stellar objects (YSOs) contain a fossil record of the recent accretion and outflow activity of their parent star-forming systems. We aim to understand whether the accretion/ejection process is similar across the entire stellar mass range of the parent YSOs.
To this end we have obtained optical to near-infrared spectra of HH~1042 and HH~1043, two newly discovered jets in the massive star-forming region RCW~36, using X-shooter on the ESO \textit{Very Large Telescope}. HH~1042 is associated with the intermediate-mass YSO 08576nr292. Over 90 emission lines are detected in the spectra of both targets. High-velocity (up to 220~km~s$^{-1}$) blue- and redshifted emission from a bipolar flow is observed in typical shock tracers. Low-velocity emission from the background cloud is detected in nebular tracers, including lines from high ionization species. 
We applied combined optical and infrared spectral diagnostic tools in order to derive the physical conditions (density, temperature, and ionization) in the jets. 
The measured mass outflow rates are $\dot{M}_{\rm jet} \sim 10^{-7}$~M$_\odot$~yr$^{-1}$. It is not possible to determine a reliable estimate for the accretion rate of the driving source of HH~1043 using optical tracers. We measure a high accretion rate for the driving source of HH~1042 ($\dot{M}_{\rm acc} \sim 10^{-6}$~M$_\odot$~yr$^{-1}$). For this system the ratio $\dot{M}_{\rm jet}/\dot{M}_{\rm acc} \sim 0.1$, which is comparable to low-mass sources and consistent with models for magneto-centrifugal jet launching.
The knotted structure and velocity spread in both jets are interpreted as fossil signatures of a variable outflow rate. While the mean velocities in both lobes of the jets are comparable, the variations in mass outflow rate and velocity in the two lobes are not symmetric. This asymmetry suggests that the launching mechanism on either side of the accretion disk is not synchronized.
For the HH~1042 jet, we have constructed an interpretative physical model with a stochastic or periodic outflow rate and a description of a ballistic flow as its constituents. We have simulated the flow and the resulting emission in position-velocity space, which is then compared to the observed kinematic structure. The knotted structure and velocity spread can be reproduced qualitatively with the model. The results of the simulation indicate that the outflow velocity varies on timescales  on the order of $100$~yr.}


\begin{document}

\maketitle

\section{Introduction}

Astrophysical jets are a ubiquitous signature of accretion. When the magnetic field of the circumstellar disk is coupled to a jet, magnetic torques can remove a significant fraction of the system's angular momentum and mass through the jet, when charged particles are ejected. Jets exist in accreting systems on various scales, ranging from young stellar objects (YSOs) up to evolved binary systems and active galactic nuclei (AGNs). They are detected in emission lines at all wavelengths, from the radio domain up to X-rays. Jets associated with optical emission are classified as Herbig-Haro (HH) objects. The velocities measured in jets are close to the escape velocity at the launch region: from a few hundred \kms in YSO jets up to relativistic velocities in AGN jets. The mass flux also scales with the mass of the central object \citep{Miley1980, Bally2007}. 

Most jets show varying velocities and shock fronts along the flow axis, which can be attributed to a launching mechanism at the jet base that is variable in time \citep{Rees1978, Raga1990}. As suggested by \citet{Reipurth1997}, the shocked structure of HH jets may reflect an FU Orionis-like accretion process: relatively short periods of intense accretion, which may be caused by thermal instabilities in the accretion disk. In principle, then, a history of the accretion is contained in the fossil record of the shock fronts in the jet. Jets thus provide the opportunity to obtain time-resolved information on the variability of the accretion process from a single observation. Given the typical spatial extent (up to a few pc) and velocities of jets it is possible to probe accretion variability on dynamical timescales up to a few thousand years, much longer than what is possible with a series of direct observations of the accretion process.

The jet launching mechanism described in the seminal paper by \citet{Blandford1982} serves as the standard model of jet formation. Matter is removed from the accretion disk by centrifugal forces, and confined along magnetic field lines that carry away the material from the disk in bipolar directions. \citet{Pudritz1983} proposed that the same mechanism drives YSO jets, as the disks of young stars are known to be threaded by magnetic fields. Most of the current understanding of protostellar jets comes from studies of low-mass objects, since these are much more numerous, and they form on longer timescales than their higher mass counterparts. However, collimated outflows are also detected around some massive YSOs \citep[e.g.][]{Torrelles2011}. As the luminosity of the central star increases, radiation pressure is thought to become a more significant driver of jets \citep{Vaidya2011}, in addition to the centrifugally driven magneto-hydrodynamic disk wind.

Just over a dozen forming (potentially) massive stars ($M \gtrsim 10$~M$_\odot$) have been reported, most of them through the detection of disk signatures at infrared, sub-mm, or radio wavelengths \citep[e.g.][]{Cesaroni2006, Zapata2009, Davies2010, Torrelles2011}. The observed rotating disks around these objects can often be associated with collimated molecular outflows. Jets from massive stars have been observed for a few objects in mm tracers, with estimated mass outflow rates in the range of $10^{-5}$ to $10^{-3}$~M$_{\odot}$~yr$^{-1}$ \citep{Cesaroni2007}. In contrast, jets from low-mass stars have been observed mostly at optical to near-infrared wavelengths, with mass outflow rates in the range of $10^{-10}$ to $10^{-6}$~M$_{\odot}$~yr$^{-1}$ \citep{Hartigan1995, Coffey2008}. The highest rates are measured in the less evolved sources, i.e. Class~I objects \citep{Hartigan1994, Bacciotti1999, Podio2006} and in FU Orionis objects \citep{Calvet1998}. The lowest rates are measured in classical T~Tauri stars (CTTS), i.e. Class~II objects. 

There currently have only been a few observations of jets from intermediate mass (2--10~M$_\odot$) YSOs in the optical to near-infrared. These are mostly evolved objects, i.e. Herbig Ae/Be stars \citep[HAeBe; e.g.][]{Levreault1984, Nisini1995, Wassell2006, Melnikov2008}, with mass outflow rates of $10^{-9}-10^{-6}$~M$_{\odot}$~yr$^{-1}$, comparable to those of CTTS. As more massive YSOs are more embedded, these are the most massive objects whose jets can be studied in the optical and near-infrared. Their observation may lead to constraints on the jet launching mechanism as it scales up to higher masses.

In this paper we study two jets which were recently discovered by \citet{Ellerbroek2011}. These authors report the discovery of a disk-jet system around the intermediate-mass ($M_* \sim 2-5$~M$_\odot$) YSO 08576nr292, located in the young massive star-forming region RCW~36 \citep[][Ellerbroek et al., in prep.]{Bik2005, Bik2006}. This region is located in the Vela molecular ridge at an estimated distance of 0.7~kpc \citep{Liseau1992}. The YSO lies at the periphery of the star-forming region, about 1$'$ ($\sim 0.2$~pc) to the West of the center of the cluster, which contains two late O-type stars. The spectrum of 08576nr292 is dominated by continuum emission from an accretion disk, with many emission lines originating in the disk, in the accretion columns and in the outflow. The jet is spatially resolved by VLT/SINFONI integral-field $H$- and $K$-band spectroscopy and shows a clumped velocity structure in [Fe\two] and H\one~emission lines. The velocity of these lines coincides with the outflow indicators in the VLT/X-shooter spectrum, suggesting that the jet originates from a spatially unresolved region close to the star. The width of the jets is not spatially resolved.

In the same study, the discovery of another jet system is reported, which emerges from object 08576nr480. Follow-up observations of both jets and their sources were carried out with X-shooter. The broad spectral coverage and intermediate spectral resolution of this instrument provide the opportunity to study the physical and kinematic properties of the jet simultaneously in the optical and in the near-infrared. The detection of optical emission in the jets resulted in their inclusion in the updated version of the catalog of Herbig Haro objects (Reipurth, private communication) under the entries HH~1042 (the 08576nr292 jet) and HH~1043 (the 08576nr480 jet). In this study we use these HH numbers when referring to the jets, and maintain the original nomenclature (08576nr292, 08576nr480) when referring to the central sources.

The analysis presented in this paper consists of three parts: a presentation of the optical to near-infrared spectra of HH~1042 and HH~1043, a description of the physical conditions in these jets, and a simulation of the kinematics of HH~1042. In Sect.~\ref{sec:obs} we describe the observations and data reduction, and in Sect.~\ref{sec:spectra} we present the obtained spectra. Physical conditions and the mass outflow rate in the jets are estimated by applying spectral diagnostics, while emission lines excited in the accretion columns provide an estimate of the accretion rate (Sect.~\ref{sec:physicalconditions}). The kinematic structure of the HH~1042 jet is simulated with an interpretative physical model; this is presented in Sect.~\ref{sec:analysis}. In Sect.~\ref{sec:disc} we discuss the appearance of the jets, the accretion and mass loss rates, and constraints on the launching mechanism. Finally, Sect.~\ref{sec:conc} contains a summary of this work.

 \begin{figure}[t]
   \centering
\includegraphics[width=0.5\textwidth]{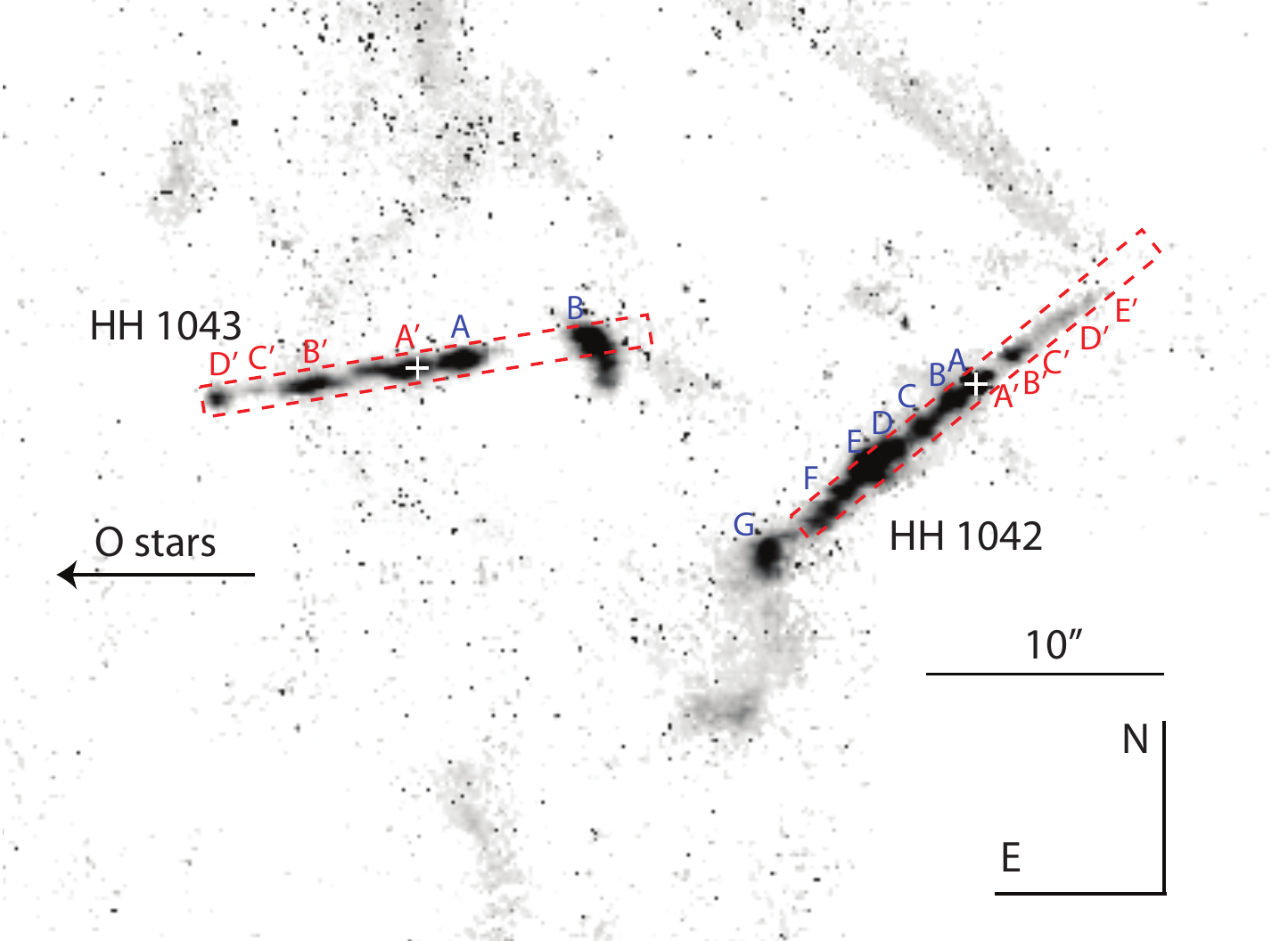}
   \caption{\small Detail of the [Fe\two] line map of RCW~36 ($d=0.7$~kpc) obtained with SINFONI (Ellerbroek et al. 2011). Merged slit positions during the X-shooter observations are indicated, as well as the positions of the knots defined in Fig.~\ref{fig:pv}. Two O stars are located in the central cluster region, $\sim 0.5'$ eastward.}
\label{fig:sfr}
    \end{figure}

\begin{table}[t]
\centering
\caption{\label{tab:obs}\normalsize{\textsc{List of the X-shooter observations.}}}
    \begin{minipage}[c]{\columnwidth}
    \renewcommand{\footnoterule}{}

\begin{tabular}{lll}
\hline
\hline
Object & HH~1042 / & HH~1043 / \\
	    & 08576nr292 & 08576nr480 \\
\hline\\[-5pt]
$\alpha$ (J2000) & 08:59:21.67 & 08:59:23.65 \\
$\delta$ (J2000) & --43:45:31.05 & --43:45:30.51 \\
Date & 19-01-2011 & 12-02-2011 \\
Exposure time\footnote{\scriptsize{Per slit position}} (s) & 600 & 600 \\
Position angle ($^\circ$, N to E) & 129 & 97 \\
Sky frame offset $\Delta \alpha, \Delta \delta$ ($\arcsec$)  & $44, 54$ & $30, 248$\\
Slit {\sc uvb/vis/nir} ($\arcsec$) & $1.0~/~0.9~/~0.4$ & $1.0~/~0.9~/~0.6$ \\
Resolution $\Delta \varv$ (\kms) & 59~/~34~/~26 & 59~/~34~/~37 \\
$K$-band seeing ($\arcsec$) & $0.6$ & $0.8$ \\
\hline
\vspace{-20pt}
\end{tabular}
\end{minipage}
\end{table}

\section{Observations and data reduction}
\label{sec:obs}

X-shooter is a cross-dispersed echelle spectrograph mounted on UT2 of the ESO \textit{Very Large Telescope}, which produces a spectrum at every spatial pixel along its 11$''$ long slit, in three separate arms: UVB (300--590 nm), VIS (550--1020~nm) and NIR (1000--2480~nm). The slit width can be chosen individually in each spectrograph arm \citep{DOdorico2006, Vernet2011}. 

The X-shooter slit was aligned with the jets, in two offset positions, both including the star, with a relative offset of $8''$ in the slit direction, so that the central 3$''$ around the source was covered twice (Fig.~\ref{fig:sfr}). In between, an exposure was taken on an empty part of the sky northeast of the target ($\Delta \alpha = +54\arcsec, \Delta \delta = +44 \arcsec $ NE of 08576nr292; $\Delta \alpha = -30\arcsec, \Delta \delta = -248 \arcsec $ SW of 08576nr480). The obtained two-dimensional spectrum covers a wavelength range from 300 to 2500~nm and $\sim19''$ along the jet, i.e. the first 9.5$''$ ($\sim$ 6,500 AU at a distance of 0.7~kpc) of the approaching (`blue') and receding (`red') lobe. In some cases the jet extends beyond the full length of the slit. Table~\ref{tab:obs} lists the characteristics of the observations.

The raw frames were reduced using the X-shooter pipeline \citep[version 1.3.7,][]{Modigliani2010}, employing the standard steps of data reduction, i.e. order extraction, flat fielding, wavelength calibration and sky subtraction, to produce two-dimensional spectra. Observations of the telluric standard star HD80055 (A0V) and the spectrophotometric standard star GD71 (a DA white dwarf) on the same night were used for the removal of telluric absorption lines and flux-calibration. The absolute flux calibration agrees to within $3-10\%$ with the existing photometry. A scaling in the relative flux calibration between the two lobes was performed on the NIR spectrum of the HH~1042 jet, where the use of a narrow slit caused some relative slitlosses between the observations of the two lobes.

The wavelength calibration in the VIS and NIR arms was refined by using the telluric OH emission lines. The UVB arm was subsequently calibrated on the VIS arm by the use of the Na\one~D feature at 589~nm, which appears in both arms. The wavelength array was then calibrated with respect to the local standard of rest (LSR). 

As shown in \citet{Ellerbroek2011} and Fig.~\ref{fig:sfr}, the objects are embedded in a star-forming region in which the ionizing stars have ionized part of the ambient cloud. Throughout the paper, we refer to this ambient interstellar medium as the `cloud'. Since photospheric features are lacking in the spectra of the central sources due to strong veiling and/or extinction, we assume the systemic velocity to be equal to the cloud velocity as measured from the nebular lines detected in our spectra (see Sect.~\ref{sec:spectra:cloud}). The nebular lines are at $-6.5 \pm 2.8$~\kms and $-1.0 \pm 6.2$~\kms ($\varv_{\rm LSR}$) for HH~1042 and HH~1043, respectively. We correct all the spectra for these values. Thus, the velocities mentioned throughout the paper and in the plots are systemic velocities ($\varv_{\rm sys}$), i.e. those with respect to the cloud velocity.

\begin{figure*}[!ht]
   \centering
\includegraphics[width=\textwidth]{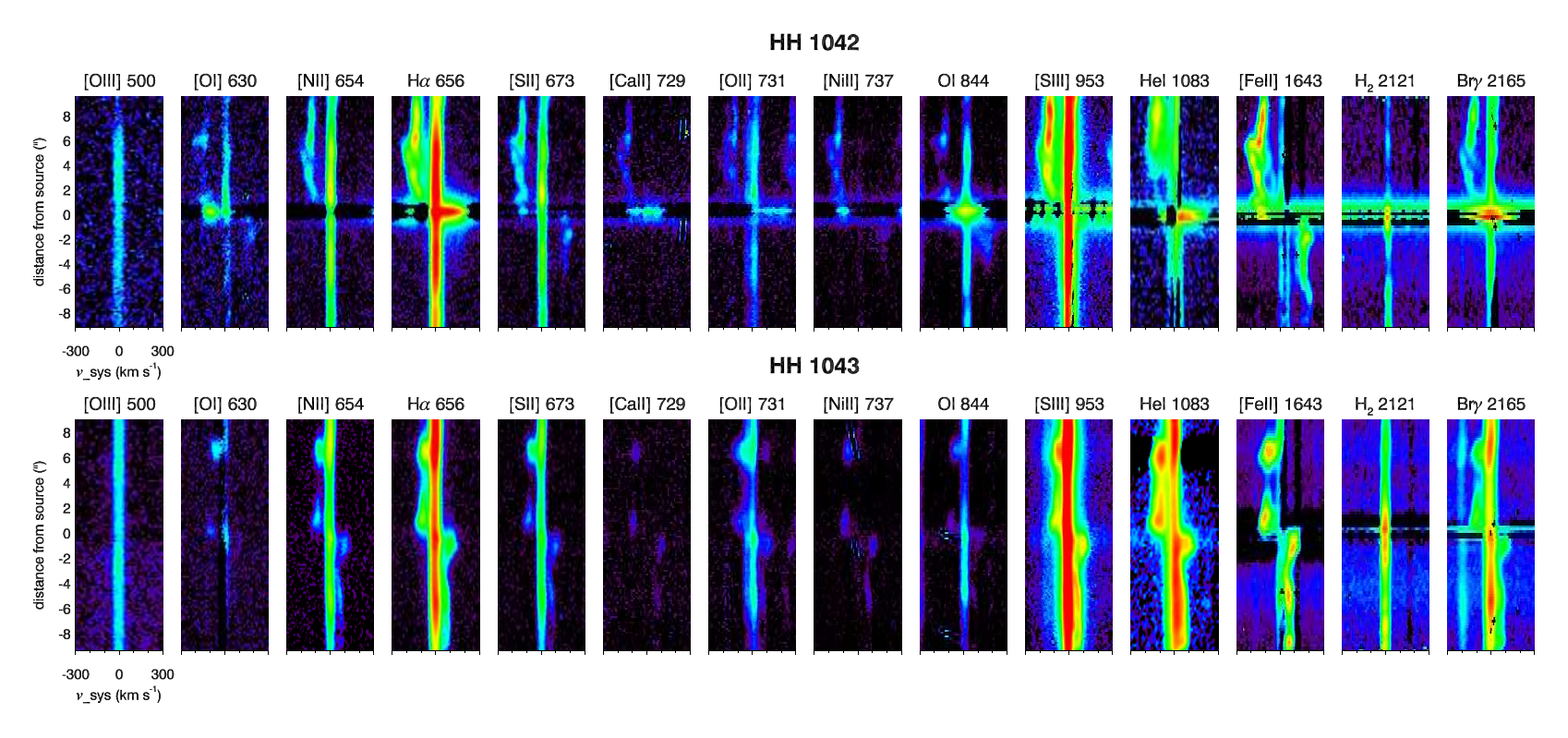}
   \caption{Position-velocity diagrams of HH~1042 (\textit{top}) and HH~1043 (\textit{bottom}) for various lines, as labeled. The absolute flux scale is logarithmic. The underlying stellar continuum (at distance = $0 ''$), where present, was subtracted using a gaussian fit. The ambient cloud produces emission in most lines at zero velocity (by definition) over the full length of the slit. In most lines the blue lobe of the HH~1042 jet is very prominent, while the red lobe suffers from extinction. The measured radial velocities in HH~1043 are significantly lower than those in HH~1042.}
\label{fig:jetspec}
    \end{figure*}

\begin{figure*}[!ht]
   \centering
\includegraphics[width=0.95\textwidth]{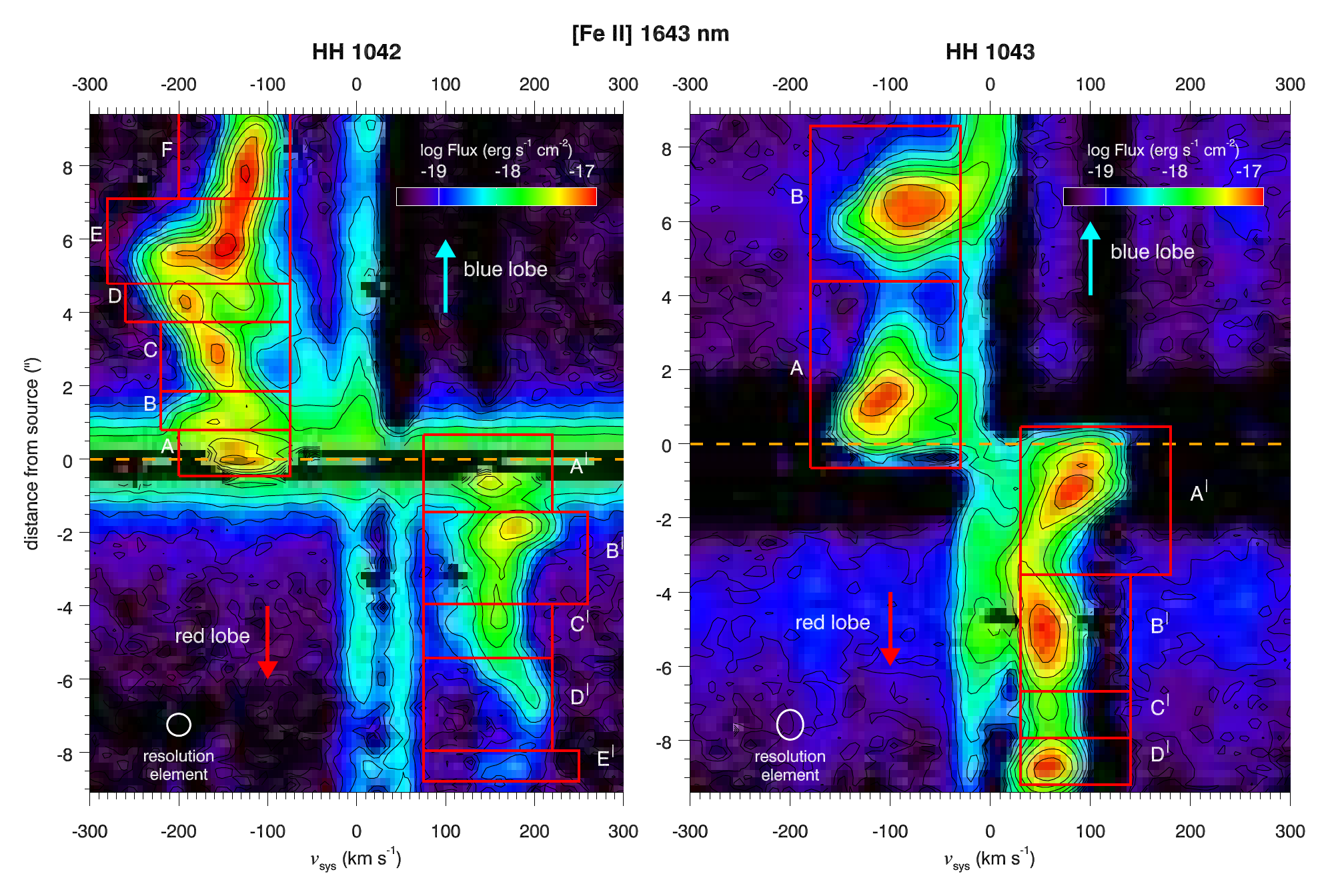}
   \caption{Position-velocity diagrams of the [Fe\two]~1643~nm line of HH~1042 (\textit{left}) and HH~1043 (\textit{right}). The underlying stellar continuum (at distance = $0 ''$) was subtracted using a gaussian fit. The positions of the knots are indicated. The dashed line indicates the position of the continuum source; 0~\kms corresponds to the systemic velocity (see text). The remnant emission between $-30$ up to $50$~\kms is a residual of the subtraction of a telluric OH emission line.}
\label{fig:pv}
    \end{figure*}

\begin{figure*}[!t]
   \centering
\includegraphics[width=\textwidth]{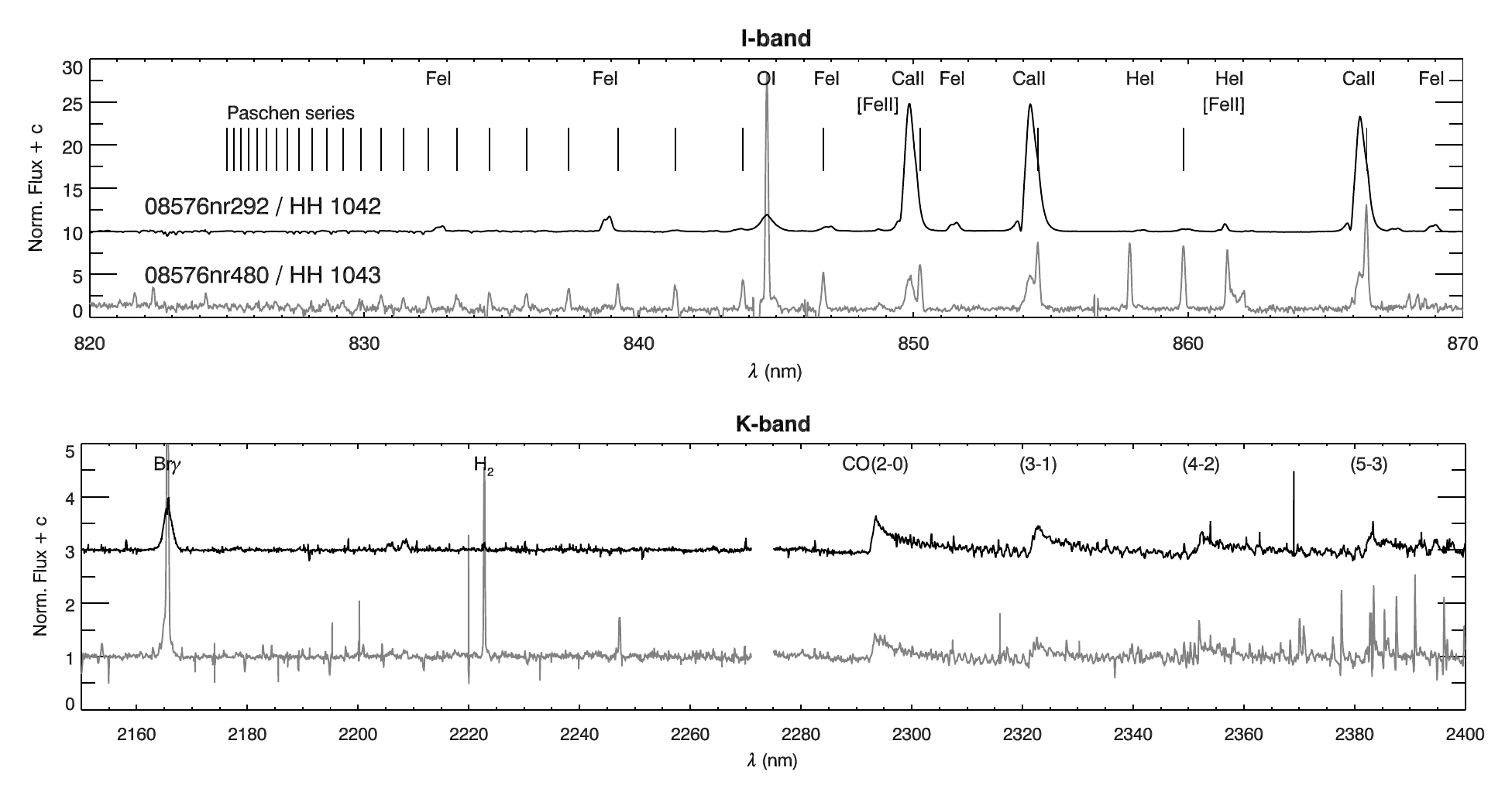}
   \caption{Sections of the continuum-normalized on-source spectra of 08576nr292 (HH~1042, upper black line) and 08576nr480 (HH~1043, lower gray line).  \textit{Top}: $I$-band; note the prominent contributions from the disk and accretion columns in 08576nr292, and the strong cloud emission in 08576nr480 (e.g. the extensive H\one~Paschen series). \textit{Bottom}: $K$-band; both sources show prominent CO bandhead emission, likely produced by a Keplerian rotating disk.}
\label{fig:onedspec}
    \end{figure*}

\begin{figure*}[!ht]
   \centering
\includegraphics[width=.9\textwidth]{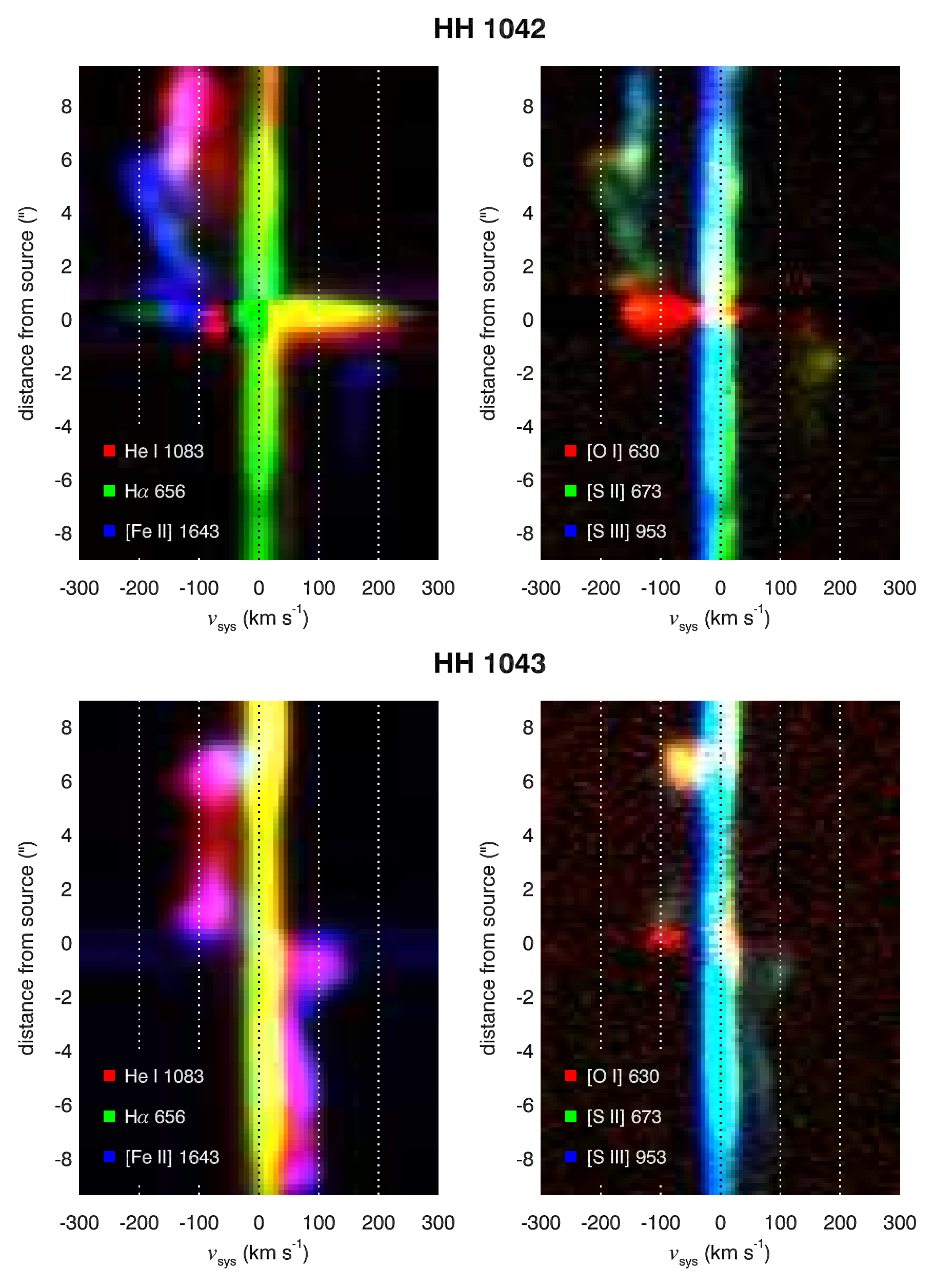}
   \caption{Merged position-velocity diagrams of three individual lines traced with colors. \textit{Top:} HH~1042; the stellar continuum of 08576nr292 (at~$0\arcsec$) was subtracted using a gaussian fit. \textit{Bottom:} HH~1043; no continuum removal was performed. Note the prominent emission of high excitation species (He\one, [S\three]) in the shock regions where the velocity drops. The [O\one] line peaks on-source, where the ejection mechanism operates, and in the shock regions.}
\label{fig:pv3color}
    \end{figure*}

Figs.~\ref{fig:jetspec}, \ref{fig:pv}, and \ref{fig:pv3color} show position-velocity diagrams of the jets in a number of emission lines. Fig.~\ref{fig:onedspec} displays the one-dimensional on-source spectra.

\renewcommand{\arraystretch}{1.2}
\begin{table*}[!ht]

\centering
\caption{\normalsize{\textsc{Identified emission lines originating in the cloud and in the jet.}}}
\begin{minipage}[l]{0.65\columnwidth}
\scriptsize{
\begin{tabular}{lllll}
\hline
\hline
$\lambda$~(nm)$^a$ & Ion &  Multiplet & Cloud & Jet \\
\hline
       372.603  &  [O~{\sc ii}] &  1F& + & --\\
       372.882  &  [O~{\sc ii}]  & 1F & + & --\\
       434.046  &  H~{\sc i} &  5--2 & + & --\\
       486.133  &  H~{\sc i} &  4--2 & + & --\\
       495.891  &  [O~{\sc iii}] &  1F& + & --\\
       500.684  &  [O~{\sc iii}] &  1F & + & --\\
       587.566  &  He~{\sc i} &  11& + & --\\
       630.030  &  [O~{\sc i}] &  1F & + & +\\
       631.206  &  [S~{\sc iii}] &  3F & + & --\\
       636.378  &  [O~{\sc i}] &  1F & + & +\\
       654.804  &  [N~{\sc ii}] &  1F & + & +\\
       656.280  &  H~{\sc i} & 3--2 & + & +\\
       658.345  &  [N~{\sc ii}] &  1F & + & +\\
       667.815  &  He~{\sc i} &  46 & + & --\\
       671.644  &  [S~{\sc ii}] &  2F & + & +\\
       673.082  &  [S~{\sc ii}] &  2F & + & +\\
       706.525  &  He~{\sc i} &  10 & + & --\\
       713.579  &  [Ar~{\sc iii}] &  1F & + & w$^b$\\
       715.516  &  [Fe~{\sc ii}] &  14F & -- & +\\
       717.200  &  [Fe~{\sc ii}] &  14F & -- & +\\
       725.445  &  O~{\sc i} &  20 & + & --\\
       728.135  &  He~{\sc i} &  45 & + & --\\
       729.147  &  [Ca~{\sc ii}] &  1F& -- & + \\
       731.992  &  [O~{\sc ii}] &  2F & + & + \\
       732.389  &  [Ca~{\sc ii}] &  1F & -- & + \\
       732.967  &  [O~{\sc ii}] &  2F & + & + \\
       733.073  &  [O~{\sc ii}] &  2F & + & + \\
       737.783  &  [Ni~{\sc ii}] &  2F & -- & + \\
       \hline
\multicolumn{5}{l}{$^a$: In air. $^b$: Very weak emission.}\\
\end{tabular}
}
\end{minipage}
\begin{minipage}[c]{0.65\columnwidth}
\scriptsize{
\begin{tabular}{lllll}
\hline
\hline
$\lambda$~(nm) & Ion &  Multiplet & Cloud & Jet  \\
\hline
       738.818  &  [Fe~{\sc ii}] &  14F & -- & + \\
       745.254  &  [Fe~{\sc ii}] &  14F & -- & + \\
       763.754  &  [Fe~{\sc ii}] &  1F & -- & + \\
       844.636  &  O~{\sc i} &  4 & + & + \\
       859.839  &  H~{\sc i} &  14--3 & + & + \\
       861.695  &  [Fe~{\sc ii}] &  13F & -- & + \\
       875.047  &  H~{\sc i} &  12--3 & + & + \\
       886.278  &  H~{\sc i} &  11--3 & + & + \\
       889.191  &  [Fe~{\sc ii}] & 13F & -- & + \\
       901.491  &  H~{\sc i} & 10--3 & + & -- \\
       905.195  &  [Fe~{\sc ii}] & 13F & -- & + \\
       906.860  &  [S~{\sc iii}] & 1F & + & + \\
       922.662  &  [Fe~{\sc ii}] & 13F & -- & + \\
       922.901  &  H~{\sc i} & 9--3 & + & + \\
       926.756  &  [Fe~{\sc ii}] & 13F & -- & + \\
       953.110  &  [S~{\sc iii}] & 1F & + & + \\
       954.597  &  H~{\sc i} & 8--3 & + & + \\
       985.026  &  [C~{\sc i}] & 1F & + & w \\
       1004.937  &  H~{\sc i} & 7--3 & + & + \\
       1028.673  &  [S~{\sc ii}] & 3F & + & + \\
       1032.049  &  [S~{\sc ii}] & 3F & + & + \\
       1033.641  &  [S~{\sc ii}] & 3F & + & + \\
       1037.049  &  [S~{\sc ii}] & 3F & + & + \\
       1083.034  &  He~{\sc i} & 1 & + & + \\
       1093.810  &  H~{\sc i} & 6--3 & + & + \\
       1188.285  &  [P~{\sc ii}] & $^3$P$_2$--$^1$D$_2$& -- & + \\
       1256.680  &  [Fe~{\sc ii}] & a$^6$D--a$^4$D & -- & + \\
       1270.347  &  [Fe~{\sc ii}] & a$^6$D--a$^4$D & -- & + \\
       \hline
       \\
\end{tabular}
}
\end{minipage}
\begin{minipage}[r]{0.65\columnwidth}
\scriptsize{
\begin{tabular}{lllll}
\hline
\hline
$\lambda$~(nm) & Ion &  Multiplet & Cloud & Jet  \\
\hline
       1278.776  &  [Fe~{\sc ii}] & a$^6$D--a$^4$D & -- & + \\
       1281.808  &  H~{\sc i} & 5--3 & + & + \\
       1294.269  &  [Fe~{\sc ii}] & a$^6$D--a$^4$D & -- & + \\
       1297.773  &  [Fe~{\sc ii}] & a$^6$D--a$^4$D & -- & + \\
       1316.485  &  O~{\sc i} & 3P--3S$^\circ$& + & + \\
       1320.554  &  [Fe~{\sc ii}] & a$^6$D--a$^4$D & -- & + \\
       1327.777  &  [Fe~{\sc ii}] & a$^6$D--a$^4$D & -- & + \\
       1533.471  &  [Fe~{\sc ii}] & a$^4$F--a$^4$D & -- & + \\
       1599.472  &  [Fe~{\sc ii}] & a$^4$F--a$^4$D & -- & + \\
       1643.549  &  [Fe~{\sc ii}] & a$^4$F--a$^4$D  & -- & + \\
       1663.766  &  [Fe~{\sc ii}] & a$^4$F--a$^4$D & -- & + \\
       1676.876  &  [Fe~{\sc ii}] & a$^4$F--a$^4$D & -- & + \\
       1680.652  &  H~{\sc i} & 11--4 & + & + \\
       1711.127  &  [Fe~{\sc ii}] & a$^4$F--a$^4$D & -- & + \\
       1736.211  &  H~{\sc i} & 10--4 & + & + \\
       1744.935  &  [Fe~{\sc ii}] & a$^4$F--a$^4$D & -- & + \\
       1797.103  &  [Fe~{\sc ii}] & a$^4$F--a$^4$D & -- & + \\
       1809.394  &  [Fe~{\sc ii}] & a$^4$F--a$^4$D & -- & + \\
       1817.412  &  H~{\sc i} & 9--4 & + & + \\
       1875.101  &  H~{\sc i} & 5--3  & + & + \\
       1895.310  &  [Fe~{\sc ii}] & a$^4$F--a$^4$D & -- & + \\
       1938.770  &  [Ni~{\sc ii}] & $^4$F--$^2$F & -- & + \\
       1944.556  &  H~{\sc i} & 8--4 & + & + \\
       2058.130  &  He~{\sc i} & $2^1$P -- $2^1$S & + & + \\
       2121.257  &  H$_2$ & 1--0 S(1) & + & -- \\
       2165.529  &  H~{\sc i} & 7--4 & + & + \\
       2222.685  &  H$_2$ & 1--0 S(0) & + & -- \\
	& & & & \\
\hline
\\
\end{tabular}
}
\end{minipage}

\label{tab:lines}
\vspace{1.cm}
\end{table*}


   \begin{figure*}[!ht]
   \centering
\includegraphics[width=.95\textwidth]{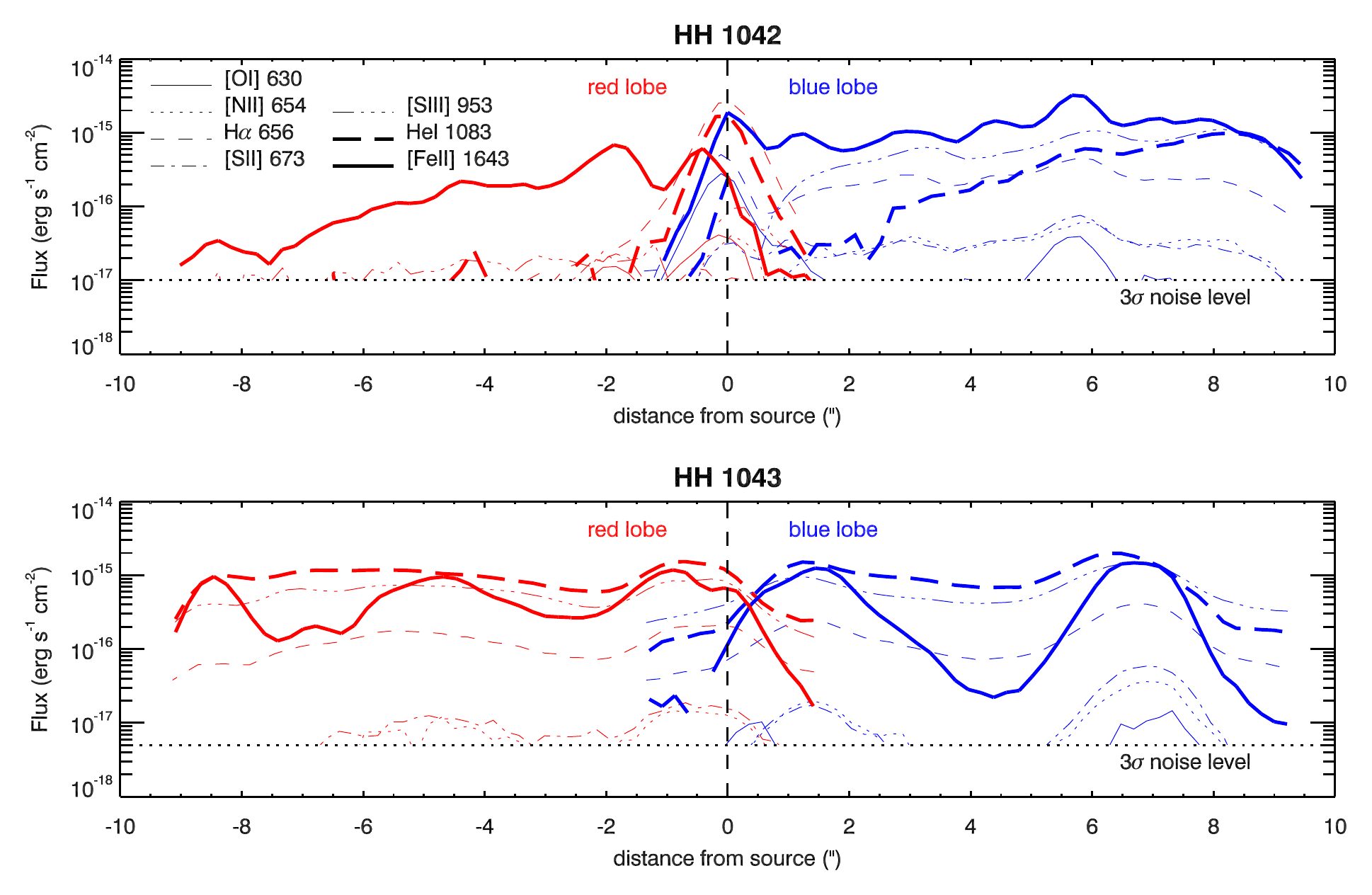}
   \caption{The observed (i.e. not corrected for extinction) integrated flux of selected lines in the jet velocity range: $-300 < \,\varv\, < -75$ \kms (blue) and $75 < \,\varv\, < 300$ \kms (red) for HH~1042; $-300 < \, \varv\, < -30$ \kms (blue) and $30 < \, \varv\, < 300$ \kms (red) for HH~1043. Note the overlap of the two slit positions; the on-source emission profiles in the blue and red lobes trace different parts of the flow. Only emission above the average $3\sigma$ noise level is shown. Note the different locations of the peaks and trends in intensity for e.g. the [Fe\two] and [He\one] lines.}
\label{fig:lineflux}
    \end{figure*}
    
\section{Analysis of the emission line spectra}
\label{sec:spectra}

The spectra obtained for the two HH objects contain more than 90 emission lines of atomic (neutral and ionized) and molecular species at different velocities. They trace various phenomena and physical conditions within the system. As we show in the following subsections the detected emission lines are originating from: (i) the ambient cloud, (ii) the circumstellar disk and the accretion/ejection region (on-source) and (iii) the jet.

\subsection{Emission from the ambient cloud}
\label{sec:spectra:cloud}

After correcting for the systemic velocity (i.e. the cloud velocity) some of the lines detected in the spectra have emission centered at $\varv_{\rm sys}=0$~\kms (see Fig.~\ref{fig:jetspec}). This emission appears in many lines (e.g., H, C, N, O and S) along the entire slit with no velocity variation, suggesting that it originates in the ambient cloud. The cloud is not emitting in the ions of refractory species that are strongly depleted in the ISM (e.g. Fe\two, Ni\two, \citealt{Sembach1996}). On the other hand, we detect cloud emission from highly ionized species (E$_{\rm ion} > 40$~keV), such as [O\three] and [Ar\three], due to the illumination by recently formed massive stars (cf. Ellerbroek et al., in prep.) in the central region of RCW~36. 

The emission is also present along the slit in the observations of \citet{Ellerbroek2011}, where the slit was placed perpendicular to the jet (see their Fig.~2). The extent of the cloud emission is confirmed by the SINFONI Br$\gamma$ and H$_2$ linemaps of RCW~36. In Tab.~\ref{tab:lines} we list the lines that are detected in the cloud (i.e. at $\varv_{\rm sys}=0$~km~s$^{-1}$), along with the lines detected in the jet (i.e., at high blue- and red-shifted velocities, see Sect.~\ref{sec:spectra:jet}).

\subsection{On-source emission: disk, accretion and outflow tracers}
\label{sec:spectra:centralsource}

Fig.~\ref{fig:onedspec} displays sections of the spectra of 08576nr292 and 08576nr480 (the central sources of HH~1042 and HH~1043, respectively), extracted from the two-dimensional frames. Neither spectrum contains photospheric absorption lines. The spectral energy distribution (SED) of 08576nr292 is dominated by emission from a circumstellar disk \citep{Ellerbroek2011}. Its spectrum is very rich in emission lines, which trace the circumstellar disk, a stellar or disk wind, (possibly) magnetospheric accretion and the onset of the jet. H$\alpha$ and the Ca\two~triplet lines show blueshifted absorption by the jet or disk wind. Although no direct signature of infall (red-shifted absorption) is detected, various emission lines associated with accretion activity (e.g. H\one, Ca\two, and He\one) are used in Sect.~\ref{sec:macc} to estimate the mass accretion rate $\dot{M}_{\rm acc}$ (Fig.~\ref{fig:macc}). The resolved double-peaked profiles of the allowed Fe\one~and Fe\two~lines in the spectrum of 08576nr292 indicate their origin in a Keplerian rotating circumstellar disk. Furthermore, the CO-bandhead feature at 2.3~$\mu$m is likely also produced in the disk. This feature is a superposition of double peaked lines. Their peak separation is determined by the mass of the central object and the inclination angle, $i$, of the system, defined as the angle between the disk rotation axis and the line of sight. It can thus be used to estimate these parameters interdependently (see Sect.~\ref{sec:disc:mass}). For a detailed description of the spectrum of 08576nr292 and a reconstruction of the system's geometry, we refer to \citet{Ellerbroek2011}.

The emission spectrum of 08576nr480 is dominated by lines from the cloud and the jet (see Sects.~\ref{sec:spectra:cloud}  and \ref{sec:spectra:centralsource}). The exceptions are a few H\one~and O~\one~lines and the Ca\two~ infrared triplet, which are thought to originate in the accretion columns and can be used to estimate the mass accretion rate; see Sect.~\ref{sec:macc}. The CO-bandhead feature at 2.3~$\mu$m is also detected. An estimate for $i$ is not obtained from this feature due to insufficient signal-to-noise and spectral resolution. 

The position of the central source on the slit was determined with a gaussian fit on the spatial profile integrated over a spectral region surrounding the emission line. The continuum emission of 08576nr480 is not detected at $\lambda < 1.5\, \mu$m. For the lines in this spectral region we have used the source position derived from the $K$-band continuum. In the plots of the jet spectra (Figs.~\ref{fig:jetspec}, \ref{fig:pv}, and \ref{fig:pv3color}) the stellar continuum (where present) was removed by subtracting the gaussian fit from the spatial profile.

\subsection{Emission from the jet}
\label{sec:spectra:jet}

In the spectra of both HH~1042 and HH~1043 we detect emission at high blue- and redshifted velocities ($100-220$~km~s$^{-1}$) in opposite directions with respect to the source position, in a number of typical jet tracers (e.g. the [O\one], [S\two] and [N\two] forbidden lines in the optical; Fig.~\ref{fig:jetspec}, Tab.~\ref{tab:lines}). This emission traces the two lobes of bipolar jets that were already apparent from the SINFONI velocity map \citep{Ellerbroek2011}. 

The emission line position-velocity diagrams (Figs.~\ref{fig:jetspec}, \ref{fig:pv} and \ref{fig:pv3color}) show a velocity structure typical of jets, with successive velocity jumps of several tens of \kms as commonly observed in shocks. The observed lines are mainly from low ionization species, as is observed in jets from low-mass stars where the typical shock velocities are $\sim 30-40$~\kms \citep[e.g.][]{Hartigan1994, Hartigan2001}. The terminal bow-shocks of jets can have much higher shock velocities, resulting in emission from high excitation species (e.g. [S\three], He\one). This is also seen in the shocks in our observations (e.g., knot E in HH~1042, Fig.~\ref{fig:pv3color}).

The three-color position-velocity diagrams (Fig.~\ref{fig:pv3color}), presented for the first time in the analysis of HH jets, highlight the variation of the excitation conditions along the jet. As usually observed in jets, strong emission is present in transitions of refractory ion species such as Fe, Ca and Ni, which are depleted onto dust grains in the ISM. When dust evaporates in the jet launching region and/or in shocks, these ion species are subsequently released in the gas phase. In the following, the main morphologic and kinematic characteristics are discussed for the two sources individually.

\subsubsection{HH~1042}
\label{sec:1042}

The emission from the bipolar jet HH~1042 in the bright  [Fe\two]~1643~nm line is shown in Figs.~\ref{fig:sfr} and \ref{fig:pv}. The blue lobe  includes seven knots labeled A to G, and extends up to 13$''$ from the central source. It is covered by the slit up to knot F, at $\sim 9''$ from the source, then it terminates in a non-collimated structure, knot G, which is visible in the SINFONI map (Fig.~\ref{fig:sfr}). The red lobe extends up to 9$''$  from the source. Beyond this point it is not detectable, most likely due to foreground extinction. The five knots in the red lobe are labeled A$'$ to E$'$ (Figs.~\ref{fig:sfr} and \ref{fig:pv}). The jet produces emission lines in allowed transitions of H, He and O, and forbidden transitions of O, P, S, Ca, Fe and Ni (Tab.~\ref{tab:lines}).

Fig.~\ref{fig:jetspec} shows the emission along the jet in a selection of lines. We see that the strongest emission is detected in the [Fe\two] lines, while also H\one, He\one~and [S\three] are prominent along the jet. The maximum brightness is at the position of the bright knot E with a flux in the [Fe\two]~1643~nm line of $2.05 \pm 0.01 \times 10^{-14}$~erg~s$^{-1}$~cm$^{-2}$. The line fluxes measured in each knot are listed in Tables~\ref{tab:flux1042blue} and \ref{tab:flux1042red} in the Appendix. The root-mean-square errors were calculated from the error spectrum (based on readout noise, flatfield, dark and bias) provided by the X-shooter pipeline.

The composite position-velocity diagrams in Fig.~\ref{fig:pv3color} show variations of the excitation conditions along the jet. [O\one] emission dominates on-source, while moving along the jet an increasing degree of ionization is seen. High ionization/excitation lines, e.g., He\one~and [S\three], dominate in the tail end, particularly knots~E and F. This could be due to the terminal shocks being the strongest in the jet.

The line flux along the jet (Fig.~\ref{fig:lineflux}) increases significantly beyond knot E in the blue lobe; in the red lobe, the emission suffers from extinction, which increases dramatically beyond 5$''$ from the source (see Sect.~\ref{sec:extinction}). The velocity in the jet is approximately 130 \kms at the base, then increases up to 220 \kms right before knot~E, after which it falls back to 140 \kms in the blue lobe (see Fig.~\ref{fig:linean}). Similarly in the red lobe, the velocities vary between 120 \kms and 210 \kms. 

Note that no clear correlation exists between the knots in the blue (A--G) and the red (A$'$--E$'$) lobes in terms of position and velocity. Asymmetries in velocity between the blue and red lobe are commonly observed in jets; this is usually attributed to an interaction with the ISM \citep[see e.g.][]{Hirth1994, Melnikov2009, Podio2011}. However, in the case of HH~1042, the average velocity is roughly equal in both lobes, although there is an uncertainty in the value of $\varv_{\rm sys}$. The variation in line flux and velocity along the jet on either side of the source is somewhat symmetric, although a definitive match between the knots in the blue and the red lobes cannot be made. We further comment on this in Sect.~\ref{sec:analysis}, where the (a)symmetry of the jet is compared with models for the outflow.

   \begin{figure*}[!t]
   \centering
\includegraphics[width=0.95\columnwidth]{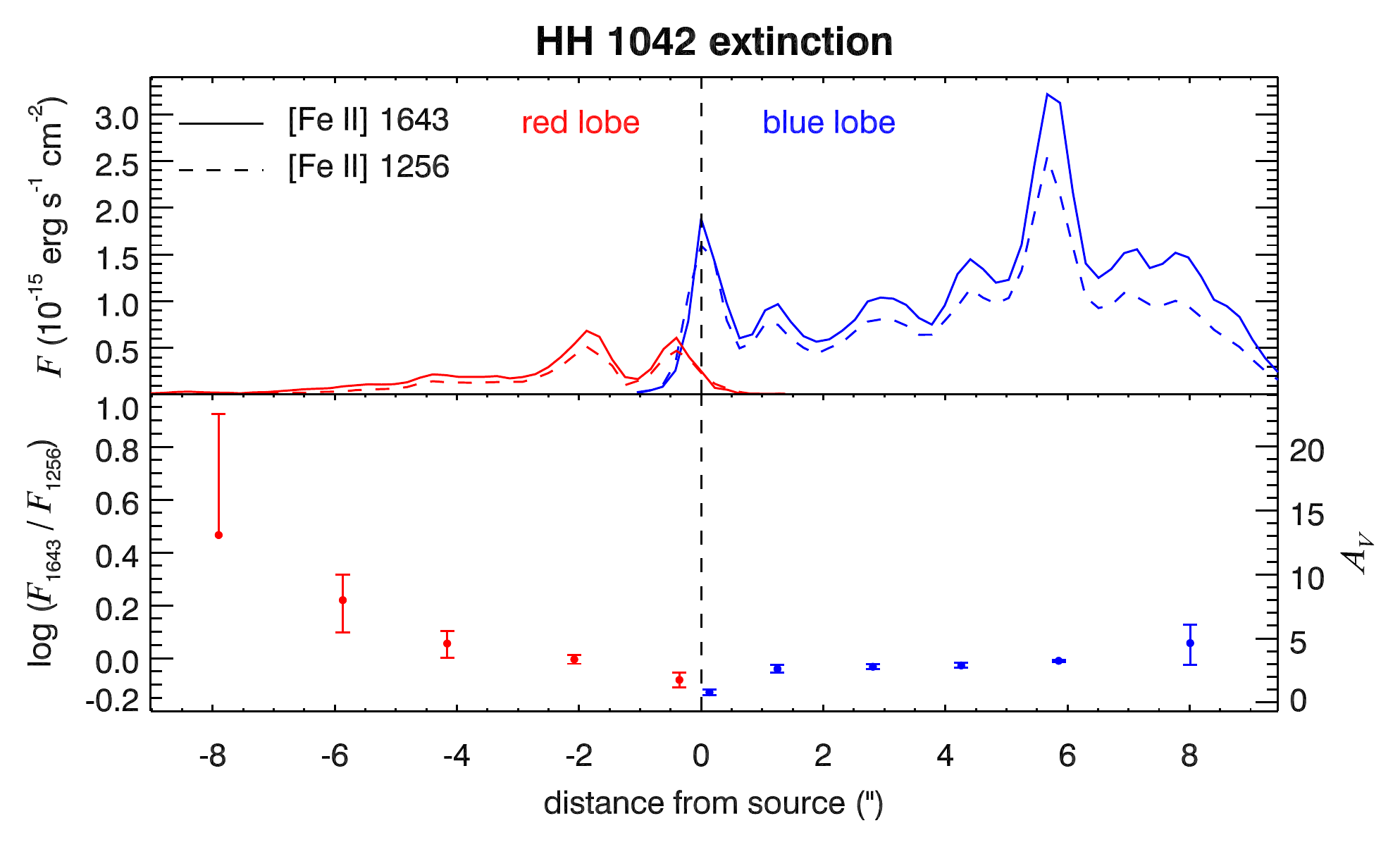}
\includegraphics[width=0.95\columnwidth]{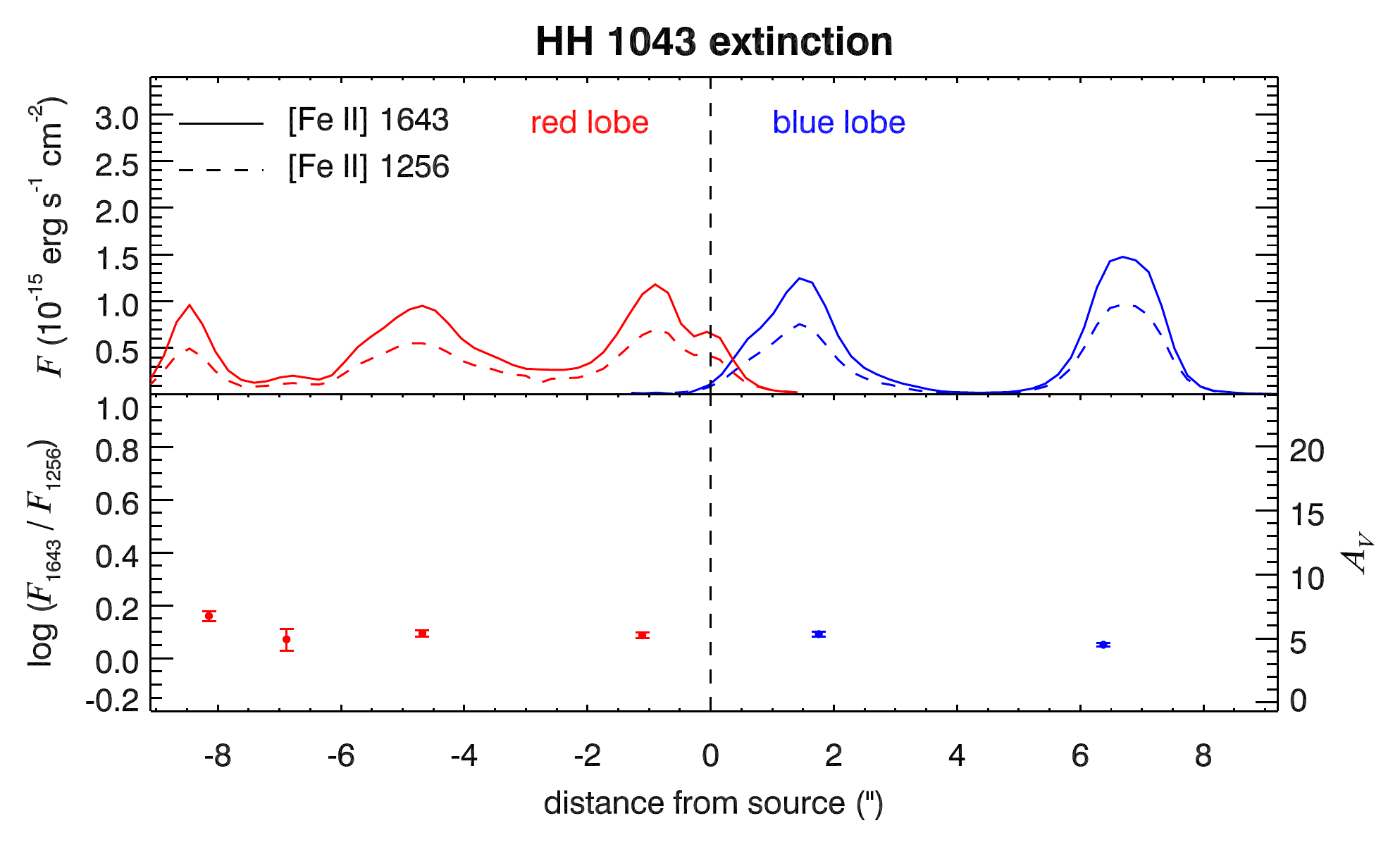}
   \caption{\textit{Top panels:} [Fe\two]~1643~nm and 1256~nm line fluxes as a function of position. \textit{Bottom panels:} [Fe\two]~1643/1256~nm flux ratio, integrated over knots. The right hand axis indicates the corresponding values of $A_V$ \citep{Cardelli1989, Quinet1996}. \textit{Left:} HH~1042; \textit{Right:} HH~1043.}
  \label{fig:av}
    \end{figure*}

\subsubsection{HH~1043}
\label{sec:1043}

The emission knots of HH~1043 are more clearly separated than those in HH~1042 (Figs.~\ref{fig:sfr} and~\ref{fig:pv}). The blue lobe consists of two knots (A and B), the latter of which ends in a bow-shock shape that is spatially resolved on the SINFONI linemap (Fig.~\ref{fig:sfr}). The red lobe is separated into four knots named from A$'$ to D$'$. The blue and red lobes both have projected lengths of $8.5''$. Note that, as in HH~1042, there is no clear symmetry between the blue and red lobes in terms of the positions of the knots. The same lines that are detected in HH~1042 are also present in HH~1043, with the addition of several H$_2$ lines and higher H\one~transitions (up to the Balmer, Paschen and Brackett jumps) due to cloud emission.

The brightest lines are again the [Fe\two], H$\alpha$, He\one~and [S\three] lines (Figs.~\ref{fig:jetspec} and \ref{fig:lineflux}). The terminal knot B has an integrated flux of $1.03 \pm 0.01 \times 10^{-15}$ \ergscm for the [Fe \two]~1643~nm line. The line fluxes measured in each knot are listed in Table~\ref{tab:flux1043} in the Appendix. Like in the other HH object, the on-source knot A has the strongest [O\one] flux, while in the terminal bow-shock (knot B), [O\one], [S\two], [S\three] and He\one~are bright (Fig.~\ref{fig:pv3color}).

The resolved bow-shock shaped feature at the end of the blue lobe and the comparable brightness of the two lobes suggest that the inclination of the jet is quite high. The average velocities in both lobes are somewhat asymmetric. Some knots in the different lobes can be matched: knots B and D$'$ are located at $\sim 8''$ either side of the source, while knots A and B$'$ are at  different distances ($2''$ and $4''$, respectively). Knot A$'$ has no visible counterpart in the blue lobe.

\section{Physical properties of the jets}
\label{sec:physicalconditions}

The detected emission lines contain information on the physical conditions of the gas in the jet. By using selected line ratios as diagnostic tools one can estimate the electron density and total density, the ionization fraction and the temperature.  In particular, the observed line ratios can be compared: (i) with those predicted by shock models \citep[e.g.][]{Hartigan1994}; or (ii) with ratios computed assuming that the employed forbidden lines are optically thin and collisionally excited, i.e. assuming that the interaction with the radiation field is negligible \citep[e.g.][]{Bacciotti1999, Podio2006}. In Table~\ref{tab:diagnostics} we summarize the line ratios used in this paper to estimate the physical properties of HH~1042 and HH~1043. Furthermore, in Sect.~\ref{sec:disc}, Fig.~\ref{fig:ragaratios}, some ratios are compared to those observed in similar sources.

In order to improve the signal-to-noise ratio, the fluxes are integrated spectrally over their profile and spatially over the defined knots. Tables~\ref{tab:flux1042blue}, \ref{tab:flux1042red} and \ref{tab:flux1043} display the integrated flux for all the emission lines in the knots where the flux exceeds the background noise by a factor 3. Table~\ref{tab:diagnostics} lists the values of the physical properties derived from selected line ratios.

\subsection{Extinction}
\label{sec:extinction}

The [Fe\two] 1643/1256~nm and 1643/1321~nm line ratios only depend on the intrinsic ratio of Einstein coefficients because the considered lines share the same upper level. Thus, the difference between observed and theoretical [Fe\two] ratios is a direct tracer of the visual extinction ($A_V$). However, the values for $A_V$ inferred from the [Fe\two] 1643/1321~nm line ratio are systematically lower than those estimated from the [Fe\two] 1643/1256~nm line ratio, because of the uncertainties affecting the computed Einstein coefficients. Moreover, when using different sets of Einstein coefficients in the literature \citep[e.g.][]{Nussbaumer1988, Quinet1996}, different values for $A_V$ are obtained. 

This issue is discussed in some detail in \citet{Nisini2005}, \citet{Podio2006} and \citet{Giannini2008}. In particular, the latter authors compared the $A_V$ values obtained from line ratios assuming different sets of Einstein coefficients with $A_V$ values estimated with other, independent methods and showed that the most reliable estimate of $A_V$ is obtained by using the [Fe\two]1643/1256~nm line ratio and the Einstein coefficients by \citet{Quinet1996}. We have adopted the same approach in this work. The [Fe\two] 1643/1256~nm line ratios and the obtained $A_V$ values are shown in Fig.~\ref{fig:av} and summarized in Table~\ref{tab:diagnostics}. A global uncertainty on these values may be caused by the value of the total-to-selective extinction $R_V$. Throughout this paper we adopt the average Galactic value of $R_V=3.1$ and the extinction law from \citet{Cardelli1989}.

In HH~1042, the red lobe appears much fainter than the blue lobe, which can be explained by the extinction trend in Fig.~\ref{fig:av}. This is consistent with the red lobe disappearing into (or behind) the molecular cloud, as was proposed in \citet{Ellerbroek2011}. Even when corrected for extinction, the flux level in the red lobe is $2-3$ times less than in the blue lobe.

The on-source value of $A_V$ measured from the [Fe\two] 1643/1256~nm ratio is much smaller ($A_V = 0.79 \pm 0.21$) than that estimated by \citet{Ellerbroek2011} from fitting the SED to a disk slope ($A_V = 8 \pm 1$). The measured on-source line ratio might be closer to unity than its true value as a result of residuals of the continuum subtraction, leading to an underestimate of the true extinction. However, even within this uncertainty, the on-source extinction would still be much lower than the value derived from the SED-fitting.  This suggests that between the jet, traced by the [Fe\two] lines, and the protostar a dusty shell or the disk might further obscure the photosphere. This phenomenon is not uncommon; for example, the extinction towards \object{DG~Tau~B} is much higher than the extinction derived from its jet emission lines \citep{Podio2011}.

In HH~1043, somewhat higher extinction values are measured in the red lobe than in the blue lobe, consistent with the slightly inclined position of the jet as discussed in Sect.~\ref{sec:1043}.
       

\begin{table*}[!ht]

\caption{\label{tab:diagnostics}\normalsize{\textsc{Physical parameters and mass loss rates estimated in the brightest knots along the HH~1042 and HH~1043 jets.}}}
\begin{minipage}[c]{\textwidth}
    \renewcommand{\footnoterule}{}
    \renewcommand{\arraystretch}{1.4}
\centering
\begin{tabular}{l l l | c c c | c c c c}
\hline
\hline
\multicolumn{3}{l}{} & \multicolumn{3}{c}{HH~1042} & \multicolumn{4}{c}{HH~1043} \\
\multicolumn{3}{l}{} & \multicolumn{2}{c}{\textit{Blue lobe}} & \multicolumn{1}{c}{\textit{Red lobe}}  & \multicolumn{2}{c}{\textit{Blue lobe}} & \multicolumn{2}{c}{\textit{Red lobe}}\\
Quantity & Diagnostic (nm) & Ref.\footnote{\scriptsize{H94: \citet{Hartigan1994}; H95: \citet{Hartigan1995}; KT95: \citet{Kwan1995}; N05: \citet{Nisini2005};  OF06: \citet{Osterbrock2006}; P06: \citet{Podio2006}; Q96: \citet{Quinet1996}}} & knot A & knot E & knot B$'$ & knot A & knot B & knot A$'$ & knot D$'$\\
\hline
$A_V$ (mag) & [Fe\two]~1643/1256 & Q96 & $  0.79^{+ 0.21}_{- 0.21}            $ & $  3.26^{+ 0.09}_{- 0.09}            $ & $  3.37^{+ 0.33}_{- 0.34}            $ & $  5.34^{+ 0.20}_{- 0.20}            $ & $  4.51^{+ 0.14}_{- 0.14}            $ & $  5.24^{+ 0.22}_{- 0.22}            $ & $  6.76^{+ 0.37}_{- 0.38} $ \\ 
$n_{\rm e}$ ($10^3$~cm$^{-3}$) & [S\two]~673/671 & OF06 & $   >8.70                $  & $  5.49^{+ 0.62}_{- 0.54} $  & $  1.54^{+ 0.32}_{- 0.27} $  & $  4.38^{+ 2.17}_{- 1.31} $  & $  6.90^{+ 1.23}_{- 0.97} $  & $  5.42^{+ 1.88}_{- 1.26} $  & \dots  \\ 
                      & [Fe\two]~1643/1533 & N05 & $  9.98^{+ 3.54}_{- 0.79} $  & $  3.51^{+ 0.23}_{- 0.23} $  & $  7.18^{+ 0.58}_{- 0.60} $  & $  8.04^{+ 0.45}_{- 0.46} $  & $  8.27^{+ 0.40}_{- 0.41} $  & $  28.6^{+ 2.0}_{- 2.0} $  & $  20.3^{+ 5.6}_{- 4.8} $  \\ 
$T_{\rm e}$ ($10^3$ K)      & [Fe\two]~1643/861 & N05 & $  10.6^{+ 0.3}_{- 0.3} $  & $  5.07^{+ 0.03}_{- 0.04} $  & $  4.78^{+ 0.11}_{- 0.11} $  & $  6.07^{+ 0.08}_{- 0.08} $  & $  5.29^{+ 0.05}_{- 0.05} $  & $  5.77^{+ 0.07}_{- 0.07} $  &   \dots   \\ 
 $x_{\rm e}$          & [N\two]~654 / [O\one]~630 & H94 & $ <0.025$    & $  \sim0.7 $                               & \dots                & \dots  & \dots  & \dots  & \dots    \\ 
 $\dot{M}_{\rm jet}$ ($10^{-8}$ M$_\odot$~yr$^{-1}$) & $n_{\rm e}$, $x_{\rm e}$ & P06 &   \dots  & $    9.59^{+ 2.02}_{- 2.08} $ & \dots  &\dots & \dots  & \dots  & \dots  \\ 
                    & $L_{\rm [S~II]}$, $n_{\rm e}$ & H95 & \dots  & $  (>0.09) $  & $  (>0.07) $  & $  (>0.05) $  & $  (>0.11) $  & $  (>0.10) $  & \dots  \\ 
                      & $L_{\rm [O~I]}$, $n_{\rm e}$ & H95 & \dots & $  (>0.03) $  & $  (>0.05) $  & $  (>0.02) $  & $  (>0.02) $  & $  (>0.02) $  & \dots  \\ 
                      & $L_{\rm [O~I]}$, $T_{\rm e}$, $\varv_{\rm shock}$ & KT95  & $  5.68^{+ 3.25}_{- 3.25} $  & $  18.8^{+ 10.7}_{- 10.7} $  & $  10.3^{+ 5.9}_{- 5.9} $  & $  32.1^{+ 18.4}_{- 18.4} $  & $ 27.1^{+ 15.5}_{- 15.5} $  & \dots  & \dots  \\ 

\hline
    \vspace{-18pt}
\end{tabular}
\end{minipage}

\end{table*}

\subsection{Electron density, temperature and ionization fraction}
\label{sec:physcond}

The values for the jet physical conditions derived from the observed line ratios are listed in Table~\ref{tab:diagnostics}. The electron density $n_{\rm e}$ is estimated using the [S\two] 673/671~nm and the [Fe\two] 1643/1533~nm line ratios. These diagnostics yield values that agree to within a factor two for most knots in the blue lobes. However, in the red lobes of both objects, the $n_{\rm e}$ values derived from [Fe\two] lines are significantly higher than those estimated from [S\two]~lines. This may be because they trace a zone of the post-shock cooling region which is located further from the shock front where the gas is more compressed, as discussed by \citet{Nisini2005} and \citet{Podio2006}. 

The electron temperature $T_{\rm e}$ is calculated from the [Fe\two] 1643/861~nm ratio, which is independent of the electron density to within an order of magnitude of $n_{\rm e}$ \citep{Nisini2005}. We detect a decreasing trend in electron temperature moving away from the sources. This is commonly observed in HH jets; see e.g. \citet{Bacciotti1999, Podio2006, Podio2011}.

Predictions by shock models \citep{Hartigan1994}, as well as theoretical line ratios demonstrate that, once $n_{\rm e}$ is determined, the [N\two]/[O\one] line ratio is almost independent from $T_{\rm e}$, hence can be used to estimate the hydrogen ionization fraction $x_{\rm e} \equiv n_{\rm H^+}/n_{\rm H}$. It should be noted that the theoretical line ratios are estimated by assuming that lines are collisionally excited and that charge exchange between O, N and H is the dominant process determining the hydrogen ionization fraction. 

We note that [O\one] emission in HH~1042 is only detected on-source and in the bright knot E, while [N\two] is detected all along the jet, but very weakly on-source (see Figs.~\ref{fig:jetspec} and \ref{fig:pv3color}). Thus, only in knot E can we can compute the [N\two]/[O\one] line ratio and derive an estimate of  $x_{\rm e}$ from the \citet{Hartigan1994} shock models. We indeed find a high ionization fraction in knot E, where the steepest velocity gradient is located, indicating that shocks may contribute to the increased ionization conditions. From the upper limit on the on-source [N\two]/[O\one] line ratios we derive $x_{\rm e} < 0.025$ in knot A. In HH~1043, the [O\one] and [N\two] emission is too weak (less than 3$\sigma$) along the whole jet, making a reliable estimate of $x_{\rm e}$ impossible. 

The [O\one] emission peaks on-source, where it is formed in the energetic (disk) wind which constitutes the base of the jet \citep{Cabrit1990}. The [O\one] velocity in knot A of HH~1042 coincides with the blueshifted absorption component of the Ca~{\sc ii} triplet lines (see Fig.~\ref{fig:onedspec}), strongly suggesting that these lines originate in the same medium. Finally, by using our $n_{\rm e}$ and $x_{\rm e}$ estimates, and assuming that the free electrons are due to the ionization of hydrogen atoms, we derive an estimate of the total density $n_{\rm H} = n_{\rm e} / x_{\rm e} = 6.46^{+1.36}_{-1.30} \times 10^3 $~cm$^{-3}$ in knot E of HH~1042.

The uncertainties on these estimated physical quantities are dominated by different effects. The uncertainty on $n_{\rm e}$ as derived from the [S\two] and [Fe\two] ratios is dominated by the error on the line fluxes, as both pairs of wavelengths are close together, making the effect of extinction negligible. The uncertainty on $T_{\rm e}$ is dominated by the error on $A_V$ as the [Fe\two] 1643~nm and 861~nm lines used for that estimate are further apart in the spectrum. The uncertainty in $x_{\rm e}$ is dependent on the errors in $A_V$ and the line flux in equal measure.

\subsection{Mass outflow rate}
\label{sec:mflux}

An important quantity in jet dynamics is the mass outflow rate, $\dot{M}_{\rm jet}$. It determines how much mass and linear momentum is injected in the surrounding cloud, and when the jet rotation is known, how much angular momentum is removed from the YSO. The ratio of the mass outflow rate to the mass accretion rate, $\dot{M}_{\rm acc}$, determines the efficiency of the star formation process. Magneto-hydrodynamic models of jets typically adopt values in the range $\dot{M}_{\rm jet}/\dot{M}_{\rm acc} \sim 0.01 - 0.1$ \citep{Konigl2000, Shu2000, Cabrit2009}. 

The mass outflow rate can be estimated from the observed line fluxes and their ratios using three different methods (see Tab.~\ref{tab:diagnostics}):
\begin{itemize}

\item[(i)] By multiplying the total density $n_{\rm H}$ with the transverse cross-section $\pi R_{\rm J}^2$ and the deprojected velocity $|\, \varv_{\rm J}\, | = \varv_\perp / \cos i$ of the jet \citep{Podio2006}:
\begin{equation}
\dot{M}_{\rm jet} =\mu \, m_{\rm H} \, n_{\rm H} \, \pi R_{\rm J}^2 \, |\, \varv_{\rm J}\, |,
\end{equation}
where $m_H$ is the proton mass and $\mu = 1.24$ was adopted for the mean atomic weight. For both objects, we adopted a value of  $R_{\rm J}=200$~AU for the jet radius. This is the average measured half width at half maximum of the spatially resolved [S\two] intensity profile in similar HH objects, which ranges from $75-300$ AU \citep[e.g.][]{Mundt1991, Reipurth2000a, Reipurth2002, Wassell2006}. The inclination is estimated at $i=17.8^{\circ+14.0}_{-2.0}$ for HH~1042 and $i=60^{\circ+15}_{-15}$ for HH~1043 (see Sect.~\ref{sec:disc:mass}). 

\item[(ii)] From the [O\one] and [S\two] emission line luminosities, if it can be assumed that all oxygen is neutral and sulphur is all singly ionized. The total line luminosity is then proportional to the number of emitting atoms in the observed volume. Adopting collisional coefficients and critical densities $n_{\rm cr}$ as in \citet[][equations A8-A10]{Hartigan1995}, we have:
\begin{equation}
\dot{M}_{\rm jet} = 9.61\times 10^{-6} \left(1 + \frac{n_{\rm cr, [OI]}}{n_{\rm e}}\right)\left(\frac{L_{\rm [OI]}}{L_\odot}\right) \left(\frac{\varv_{\rm J}}{l_{\rm knot}}\right) \, {\rm M_\odot yr^{-1}}
\end{equation}
and
\begin{equation}
\dot{M}_{\rm jet} = 1.43\times 10^{-3} \left(1 + \frac{n_{\rm cr, [SII]}}{n_{\rm e}}\right)\left(\frac{L_{\rm [SII]}}{L_\odot}\right) \left(\frac{\varv_{\rm J}}{l_{\rm knot}}\right) \, {\rm M_\odot yr^{-1}}.
\end{equation}

\item[(iii)] From the [O\one] line luminosity if this is produced by post-shock cooling, by using Eq.~(A14) from \citet{Hartigan1995}, which is adopted from \citet{Kwan1995}: 
\begin{equation}
\dot{M}_{\rm jet} = \frac{\varv_{\rm J}}{\varv_{\rm sh}}\,  \frac{f \, \mu \, m_{\rm H} \, L_{\rm [O~I]}}{\frac{3}{2}\, kT_{\rm e}}.
\end{equation}
We are unable to determine the shock velocity $\varv_{\rm sh}$ from spatially resolved shock fronts. Typical values from the models discussed in \citet{Hartigan1994} are in the range $\varv_{\rm jet}/\varv_{\rm sh} \sim 5-10$; we adopt $\varv_{\rm jet}/\varv_{\rm sh}=10$. It is assumed that a fraction $1/f$ of the total luminosity radiated below 6000~K is in the [O\one]~630~nm line. As in \citet{Kwan1995}, we adopt $f=3.5$.
\end{itemize}

The uncertainties on the $\dot{M}_{\rm jet}$ estimates from method (i) are mainly due to the assumption of the jet radius and the uncertainty on the inclination angle, while for method (ii) and (iii), which are based on the luminosities of forbidden lines, the main source of error is the uncertainty on the extinction and the distance to the source and -- for method (iii) -- the estimated shock velocity. An uncertainty in the absolute flux calibration due to the correction for slit losses may be present, although this marginally contributes to the error budget.  The excitation models used in method (ii) assume that all oxygen is neutral and all sulphur is singly ionized. As emission in [O\two] and [S\three] lines is detected in the jets, the estimates for $\dot{M}_{\rm jet}$ obtained from this method are considered lower limits. The high excitation conditions in the jet may point to an external source of radiation (see Sect.~\ref{sec:disc:physicalproperties}). However, since  even higher ionization species are detected in the cloud and not in the jet, any external radiation field causing this emission is not expected to affect the estimates on the physical conditions and mass outflow rates in a significant way.


Method (i) can only be applied in knot E of HH 1042, because it is the only knot where we can retrieve an estimate of $x_{\rm e}$. Method (iii) can be applied in those knots where [O\one] is detected and we have an estimate of $A_V$. On-source, where the gas is thought to be almost neutral as indicated by the derived upper limit on $x_{\rm e}$ for HH 1042 (see Tab.~\ref{tab:diagnostics}), we cannot apply this method because we do not have a reliable estimate of $n_{\rm e}$, due to the faintness of the [S\two] lines.

Tab.~\ref{tab:diagnostics} summarizes the values and lower limits obtained by applying the explained methods. The absolute values of $\dot{M}_{\rm jet}$ obtained by methods (i) and (iii) agree well and indicate $\dot{M}_{\rm jet} \sim 10^{-7}$~M$_\odot$~yr$^{-1}$ in both HH~1042 and HH~1043 (see Tabs.~\ref{tab:diagnostics}~and~\ref{tab:macc-mjet}). In both jets, no significant asymmetry is found between the red and blue lobe values of $\dot{M}_{\rm jet}$.

   \begin{figure}[ht]
   \centering
\includegraphics[width=\columnwidth]{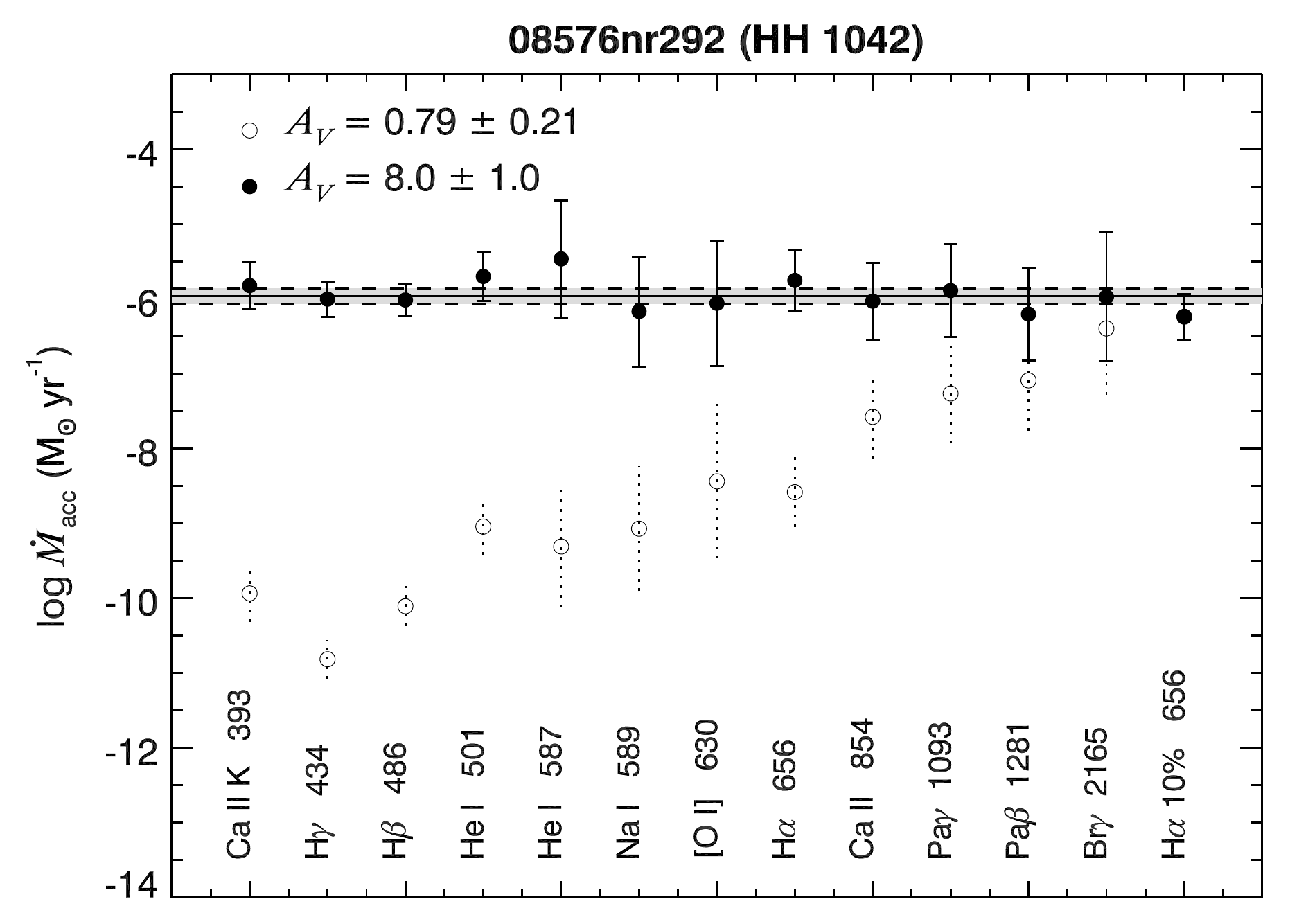}
   \caption{Values for $\dot{M}_{\rm acc}$ of 08576nr292 derived with A$_V=0.79 \pm 0.21$ (open circles) and A$_V=8 \pm 1$ (closed circles).}
  \label{fig:macc}
    \end{figure}


\begin{table}[!h]
\centering
\caption{\label{tab:macc}\normalsize{\textsc{Estimated mass accretion rate.}}}
\begin{minipage}[c]{\columnwidth}
    \renewcommand{\footnoterule}{}

\scriptsize{
\begin{tabular}{l c c c c}
\hline
\hline
Line, $\lambda$ & $\log$ Flux \footnote{\scriptsize{The errors are less than 0.01 dex.}} & $\log$ Flux  & $\log \dot{M}_{acc}$ & Reference\footnote{\scriptsize{G08: \citet{Gatti2008}; HH08: \citet{Herczeg2008}; M98: \citet{Muzerolle1998b}; N04: \citet{Natta2004}}}\\
  (nm) &  (erg s$^{-1}$ cm$^{-2}$) & (erg s$^{-1}$ cm$^{-2}$) & (M$_\odot$ yr$^{-1}$) & \\
\hline
\multicolumn{5}{c}{08576nr292 (HH~1042)}\\
  & $A_V=0$ & $A_V=8 \pm 1$ & $A_V=8 \pm 1$ & \\

\hline
Ca\two~K~393 & $ -15.28 $ & $ -10.66 \pm  0.04 $ & $  -5.95 \pm  0.30 $ & HH08 \\
H$\gamma$~434 & $ -15.71 $ & $ -11.43 \pm  0.06 $ & $  -6.03 \pm  0.21 $ & HH08 \\
H$\beta$~486 & $ -14.83 $ & $ -11.05 \pm  0.04 $ & $  -5.95 \pm  0.21 $ & HH08 \\
He\one~501 & $ -15.21 $ & $ -11.59 \pm  0.06 $ & $  -5.66 \pm  0.31 $ & HH08 \\
He\one~587 & $ -15.35 $ & $ -12.46 \pm  0.17 $ & $  -5.51 \pm  0.74 $ & HH08 \\
Na\one~D~589 & $ -14.96 $ & $ -12.03 \pm  0.07 $ & $  -6.20 \pm  0.73 $ & HH08 \\
$[$O\one$]$~630 & $ -14.36 $ & $ -11.64 \pm  0.04 $ & $  -6.09 \pm  0.84 $ & HH08 \\
H$\alpha$~656 & $ -13.02 $ & $ -10.42 \pm  0.01 $ & $  -5.77 \pm  0.40 $ & HH08 \\
Ca\two~854 & $ -12.89 $ & $ -11.19 \pm  0.01 $ & $  -6.02 \pm  0.51 $ & HH08 \\
Pa$\gamma$~1093 & $ -13.14 $ & $ -12.02 \pm  0.02 $ & $  -5.89 \pm  0.62 $ & G08 \\
Pa$\beta$~1281 & $ -12.68 $ & $ -11.82 \pm  0.01 $ & $  -6.20 \pm  0.62 $ & M98 \\
Br$\gamma$~2165 & $ -12.87 $ & $ -12.50 \pm  0.01 $ & $  -5.97 \pm  0.86 $ & M98 \\
H$\alpha$ 10\% width &  686 km~s$^{-1}$ &  \dots & $  -6.24 \pm  0.30 $ & N04 \\
\hline
\multicolumn{5}{c}{08576nr480 (HH~1043)}\\
  & $A_V=0$ & $A_V=12 \pm 3$ & $A_V=12 \pm 3$ & \\
\hline
Ca\two~854 & $ -15.66 $ & $ -13.11 \pm  0.09 $ & $  -7.97 \pm  0.60 $ & HH08 \\
Pa$\gamma$~1093 & $ -14.32 $ & $ -12.64 \pm  0.05 $ & $  -6.73 \pm  0.65 $ & 
G08 \\
Pa$\beta$~1281 & $ -13.82 $ & $ -12.52 \pm  0.03 $ & $  -7.00 \pm  0.67 $ & 
M98 \\
Br$\gamma$~2165 & $ -13.96 $ & $ -13.40 \pm  0.03 $ & $  -7.11 \pm  0.93 $ & 
M98 \\

\hline
    \vspace{-18pt}
\end{tabular}
}
\end{minipage}

\end{table}

\begin{table}[!h]
\caption{\label{tab:macc-mjet}\normalsize{\textsc{Mass outflow and accretion rate.}}}
    \begin{minipage}[c]{\columnwidth}
    \renewcommand{\footnoterule}{}

\centering
\begin{tabular}{lll}
\hline
\hline
Object & HH~1042 / & HH~1043 / \\
	    & 08576nr292 & 08576nr480 \\
\hline\\[-5pt]
$\dot{M}_{\rm jet}$ (M$_\odot$ yr$^{-1}$) & $8.86^{+1.63}_{-1.66} \times 10^{-8} $ & $2.92^{+1.19}_{-1.19} \times 10^{-7}$ \\

$\dot{M}_{\rm acc}$  (M$_\odot$ yr$^{-1}$) & $1.10^{+0.29}_{-0.21}\times10^{-6}$ &$5.50^{+6.53}_{-2.99}\times10^{-8}$ \\
\hline
\vspace{-20pt}
\end{tabular}
\end{minipage}
\end{table}

\subsection{Accretion rate}
\label{sec:macc}

The accretion rate, $\dot{M}_{\rm acc}$, is derived from the accretion luminosity, $L_{\rm acc}$, which can be determined by measuring the UV excess flux emitted by the accretion flow close to the star \citep[e.g.][]{Hartigan1995, Gullbring1998}. Radiative transfer models and spectra show that certain spectral lines are formed in the accretion flow \citep{Hartmann1994}. Therefore one would expect that the line strength correlates with the accretion luminosity, which was confirmed for CTTS in subsequent studies \citep[e.g.][]{Muzerolle1998b, Muzerolle1998a}. This correlation between line strength and $L_{\rm acc}$ is consistent across the mass spectrum, from brown dwarfs up to HAe stars (e.g., \citealt{Natta2004, Mendigutia2011}, see also Fig.~\ref{fig:macc-mjet}). 

The mass accretion rate can be related to the accretion luminosity, following \citet{Gullbring1998}. The accretion luminosity is equal to the amount of energy per unit time released from the gravitational field when material falls onto the stellar surface -- along magnetic field lines -- from the radius $R_{\rm in}$ where the disk is truncated by the stellar magnetic field:
\begin{equation}
\dot{M}_{\rm acc} =\left(1- \frac{R_*}{R_{\rm in}} \right)^{-1} \frac{L_{\rm acc}R_*}{GM_*}.
\end{equation} 
We adopt $M_*/R_* \sim 1$ (in solar units) for the central star, consistent with pre-main sequence models \citep[PMS,][]{Siess2000}, and assume a typical value of $R_{\rm in} \sim 5\,R_*$ \citep{Shu1994}. 

Fig.~\ref{fig:macc} shows that for 08576nr292, adopting an extinction $A_V=0.79 \pm 0.21$ found from the [Fe\two] line ratio, the accretion diagnostics are not consistent. Instead, using $A_V=8.0\pm1.0$ as estimated from SED fitting by \citet{Ellerbroek2011}, we obtain consistent results for tracers across the entire spectral range: $\log L_{\rm acc}/L_\odot = 1.53 \pm 0.10$, which is of the same order as the stellar luminosity (Sect.~\ref{sec:disc}). Subsequently $\log \dot{M}_{\rm acc}=-5.96 \pm 0.10$ M$_\odot$ yr$^{-1}$. This confirms that the on-source extinction is much higher than that estimated at the base of the jet from [Fe\two] lines.

The spectrum of the driving source of HH~1043, 08576nr480, only exhibits a few accretion tracers (Ca\two, Br$\gamma$, Pa$\beta$ and Pa$\gamma$). Using these tracers, the accretion luminosity is found to be $\log L_{\rm acc}/L_\odot = 0.23 \pm 0.34$, and $\log \dot{M}_{\rm acc}=-7.26 \pm 0.34$ M$_\odot$ yr$^{-1}$. We adopted $A_V=12\pm3$ from the SED-fitting method described in \citet{Ellerbroek2011}. Tab.~\ref{tab:macc-mjet} lists the derived mass outflow and accretion rates. 
  
\section{Kinematics analysis}
\label{sec:analysis}

In this section, we present our analysis of the kinematics of the HH~1042 jet, as well as an interpretative model that we use to simulate the recent outflow history of the jet. Our aim is to learn more about the central YSO and the jet launching mechanism from the fossil record of the outflow history contained in the jet. We focus on the position-velocity diagram of the [Fe\two]~1643~nm line. This line is one of the brightest in the spectra and traces the largest velocity range. An additional advantage is that the detector resolution is the highest in the near-infrared arm ($\Delta \varv \sim 26$~km~s$^{-1}$, see Table~\ref{tab:obs}). Note, however, that given the spatial and spectral resolution of the observations, and the lack of proper motion measurements, one cannot formulate a unique model that reproduces the data. Instead, in this section we simulate the general shape of the emission pattern, derive timescales relevant for the ejection mechanism, and draw qualitative conclusions about the physics within the flow.

\begin{figure*}[!ht]
   \centering
\includegraphics[width=\textwidth]{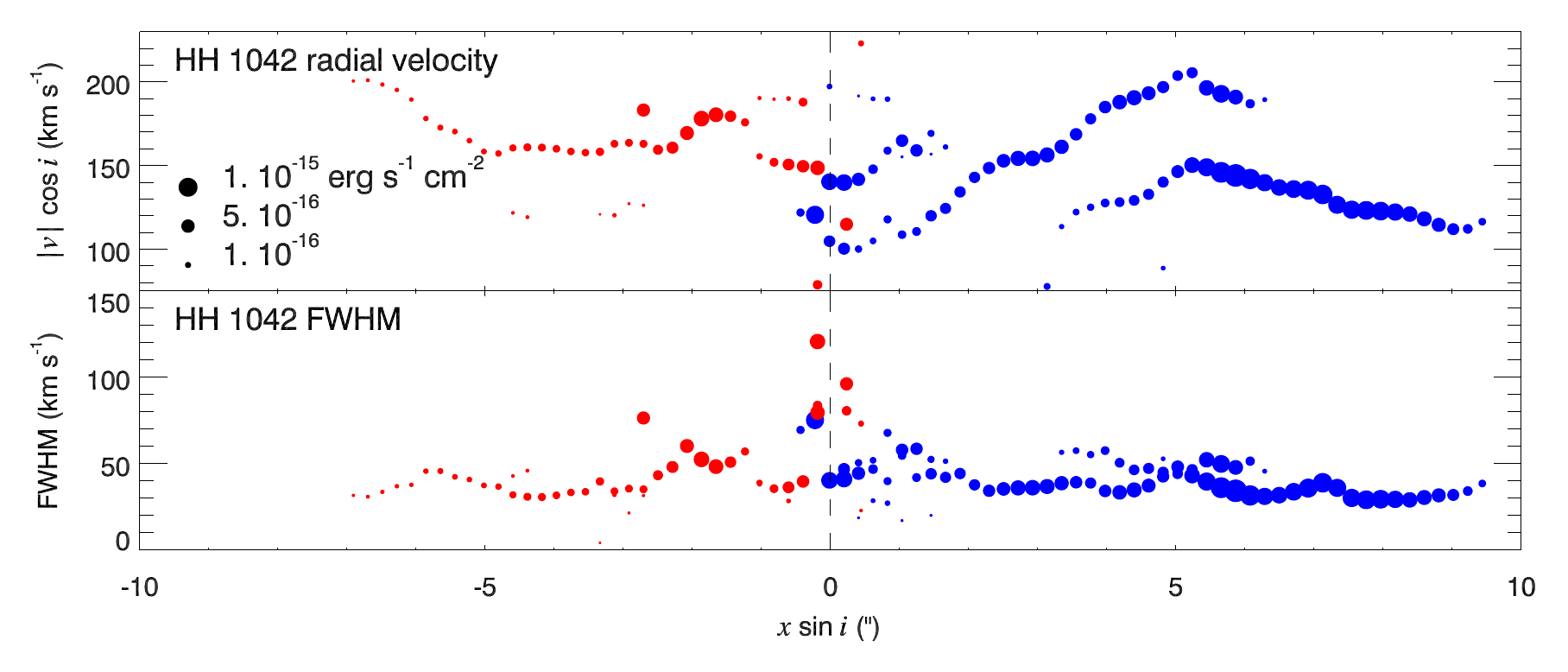}
\includegraphics[width=\textwidth]{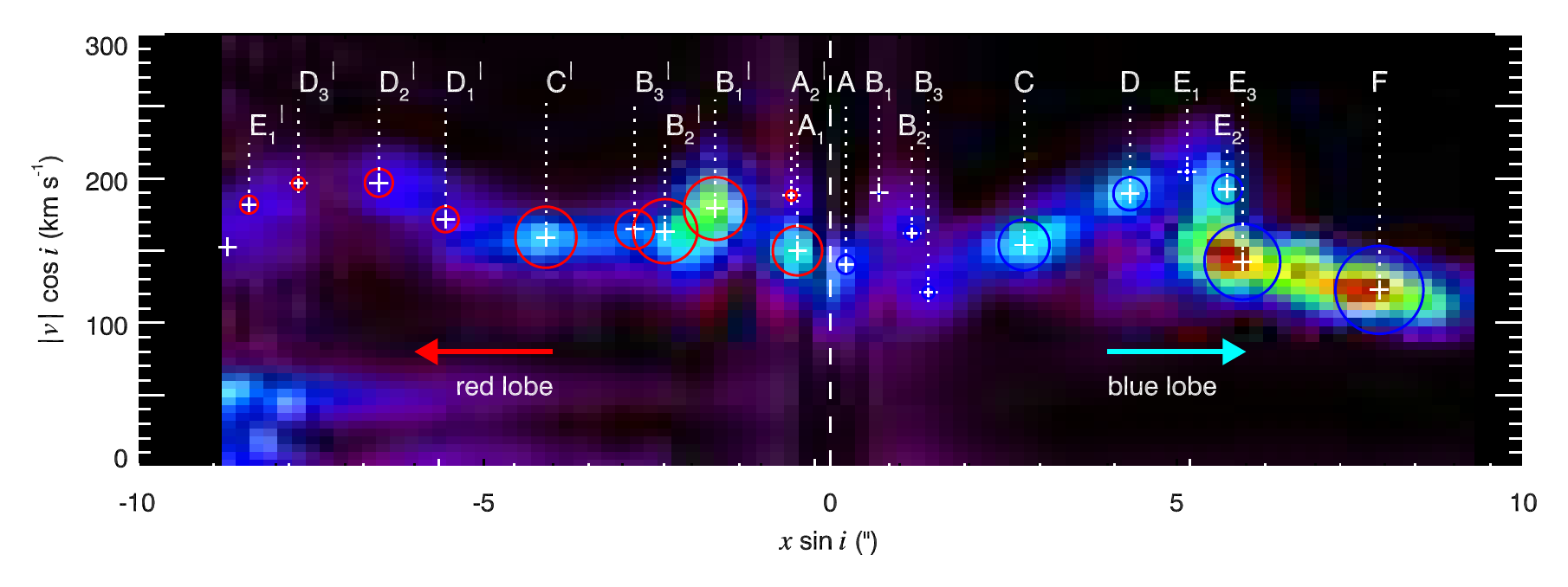}
   \caption{\textit{Top:} Peak radial velocity, $|\, \varv\, | \cos i$, and FWHM of the velocity components detected along the HH~1042 jet, obtained by means of a multiple gaussian fit of the [Fe\two]~1643~nm line velocity profile at each spatial pixel (0.2$''$) along the jet. The size of the symbols represents the integrated flux of each velocity component. \textit{Bottom:} Position-velocity diagram of the [Fe\two]~1643~nm line in HH~1042. The $y$-axis corresponds to the absolute value of the radial velocity, in order to show the (a)symmetry between the two lobes. The flux is corrected for extinction and the red lobe (\textit{left}) is enhanced by a factor of 3, in order to compare the emission patterns of both lobes. The source is located at 0$''$ in position space. The letters indicate the knots which were integrated in flux for the kinematics analysis. The size of the circles corresponds to the integrated flux of the knots.}
\label{fig:linean}
    \end{figure*}

The position-velocity diagram of the [Fe\two] line shows an outflow variable in velocity (Figs.~\ref{fig:pv} and \ref{fig:linean}). This kinematic structure is assumed to be the result of an outflow which varies in both velocity and mass outflow rate at the launch site at the base of the jet. The outflow rate is reconstructed by comparing the data to a ballistic outflow model that assumes a launch mechanism that is either stochastic or periodic. The simulation consists of two ingredients: a characterization of the temporal variation of the mass outflow rate at the base of the jet (the `input physics') and a description of the flow of material through the jet (the `interaction physics'). The former is reconstructed in Sect.~\ref{sec:analysis:measure} for both the stochastic and the periodic mechanisms; the latter is explained in Sect.~\ref{sec:analysis:model}. The results of the simulations are presented in Sect.~\ref{sec:analysis:simulationresults}. The kinematics of HH~1043 are not simulated, as there are too few emission knots along the jet to constrain the model parameters.

   \subsection{`Input physics': outflow rates $\varv(t)$, $\dot{m}(t)$}
\label{sec:analysis:measure}

Time variability is a known property of accretion-ejection mechanisms and has been measured in both the accretion luminosity \citep[e.g.][]{Herbst1994, Alencar2002, Hillenbrand2012} and outflow activity \citep[e.g.][]{Micono1998} of YSOs on timescales ranging from days to years. It has been argued by \citet{Hartmann1985} that this reflects an intrinsic variability of the accretion process. One may therefore expect all observables correlated with accretion, i.e. the luminosity, mass accretion and outflow rates as well as outflow velocities, to be variable in time. The characteristics of these variations are not well constrained, but they are expected to be either purely periodic (e.g. due to disk rotation or binary interaction), quasi-periodic (e.g. due to the interplay between magnetic stress and pressure in the accretion disk), or stochastic (e.g. due to chaotic processes that depend on many physical parameters). Different periodicities and timescales, tracing different mechanisms, may exist within one accretion system.

In principle, both the outflow velocity, $\varv(t)$, and the mass outflow rate, $\dot{m}(t)$ can be variable. A varying $\varv(t)$ results in differences in velocity and line flux (because of the formation of shocks) along the jet; a varying $\dot{m}(t)$ introduces a variation in density and hence line flux along the jet. In this section, we explore how well we can reconstruct the outflow velocity and mass outflow rate from the data. 

We estimate the launch time of the material along the jet from its present position and velocity. From the position-velocity diagram of the [Fe\two]~1643~nm line, a one-dimensional spectral profile is extracted at every pixel of width $=0.2''$ along the jet. To this emission profile, a superposition of one or more one-dimensional gaussian functions is fitted. We have made use of the IRAF routine \verb splot, which deblends multiple gaussian components and calculates errors on the fit parameters based on a Poisson noise model. The number of components is increased until adding another component does not significantly improve the fit; in most cases this amounts to one or two components per row. For every spatial pixel row of every emission line, this results in a list of emission components, their velocities, widths, and fluxes (Fig.~\ref{fig:linean}, top). 

To increase the signal-to-noise ratio, these emission components are grouped into knots in position-velocity space (Fig.~\ref{fig:linean}, bottom). The knots are similar to those defined in Sect.~\ref{sec:spectra}; some are split into several components with different radial velocities (labeled with subscripts, e.g. E$_1$, E$_2$). The velocities and positions of the knots are determined by averaging them over the values of the constituent components, where the velocities are weighted by their inverse squared errors. 

\begin{figure}
   \centering
\includegraphics[width=\columnwidth]{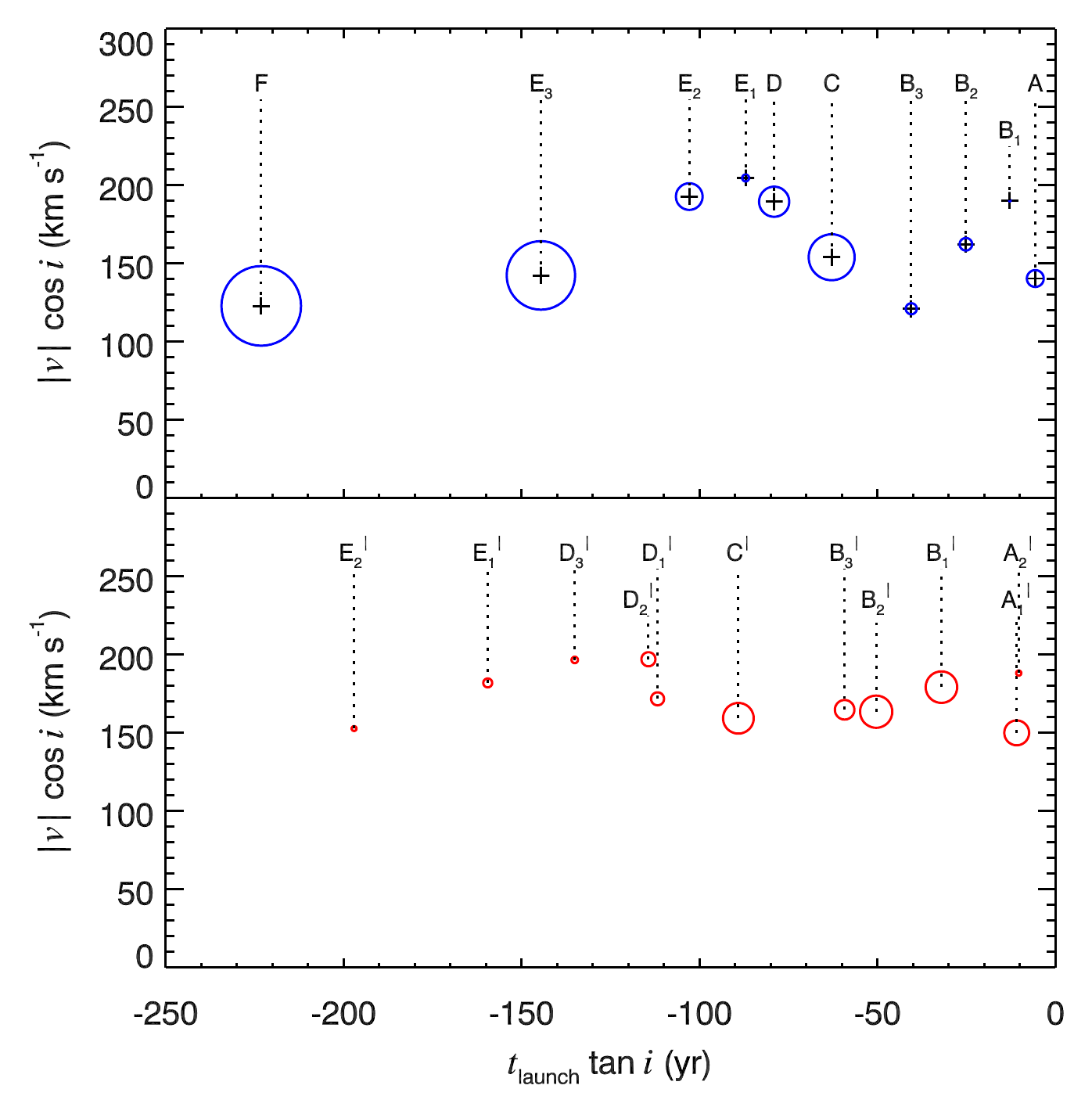}
   \caption{Radial velocity, $|\, \varv \,| \cos i$, versus launch time, $t_{\rm launch} \tan i$, assuming no collisions along the flow. Symbols represent the knots in the blue lobe (\textit{above}, blue circles) and red lobe (\textit{below}, red circles) as defined in Fig.~\ref{fig:linean} (bottom). The area of a circle corresponds to the line flux integrated over the knot.}
\label{fig:tfit}
    \end{figure}

We estimate for every knot with measured line-of-sight velocity, $\varv \cos i$, and projected position in the plane of the sky, $x \sin i$, the launch time, by assuming that no collisions occur in the flow (defined as the case $\hat{\eta}=-1$, see next section):
\begin{equation}
t_{\rm launch} \tan i= -\frac{x \sin i}{\varv \cos i}
\label{eq:tlaunch}
\end{equation}
The results are displayed in Fig.~\ref{fig:tfit}. The intervals between the knots and the velocity variation appear to be somewhat regular, particularly in the blue lobe. One might suspect that this reflects a purely periodic outflow rate, with single-mode sine waves for $\varv(t)$ and $\dot{m}(t)$, which would be the simplest case of an outflow rate which is variable in time. However, apparent regularities in time series may wrongly suggest periodicity to the human eye. A stochastic or quasi-periodic outflow rate may also produce an apparently periodic signal in the outflow pattern. We therefore consider two possibilities for the variations in $\varv(t)$ and $\dot{m}(t)$: a stochastic variation, the parameters of which are obtained by Fourier analysis; and a purely periodic variation, the parameters of which are obtained by an iterative, direct fitting procedure. In both cases, a different periodicity is allowed for $\varv(t)$ and $\dot{m}(t)$. From this point on we refer to these two cases of simulated outflow rates as `stochastic model' and `periodic model'.

   \begin{figure}[!t]
   \centering
\includegraphics[width=0.9\columnwidth]{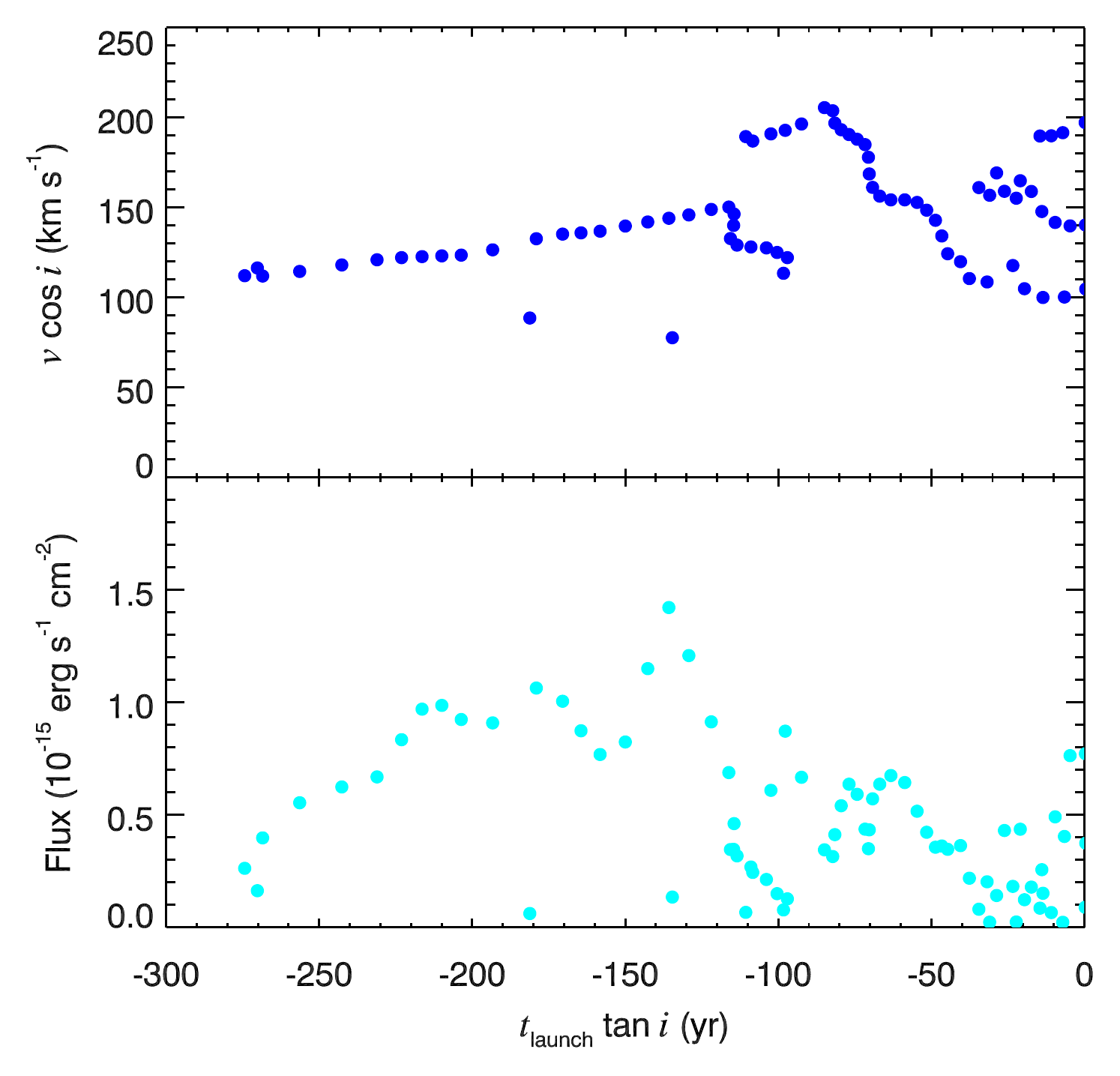}\\
\includegraphics[width=0.9\columnwidth]{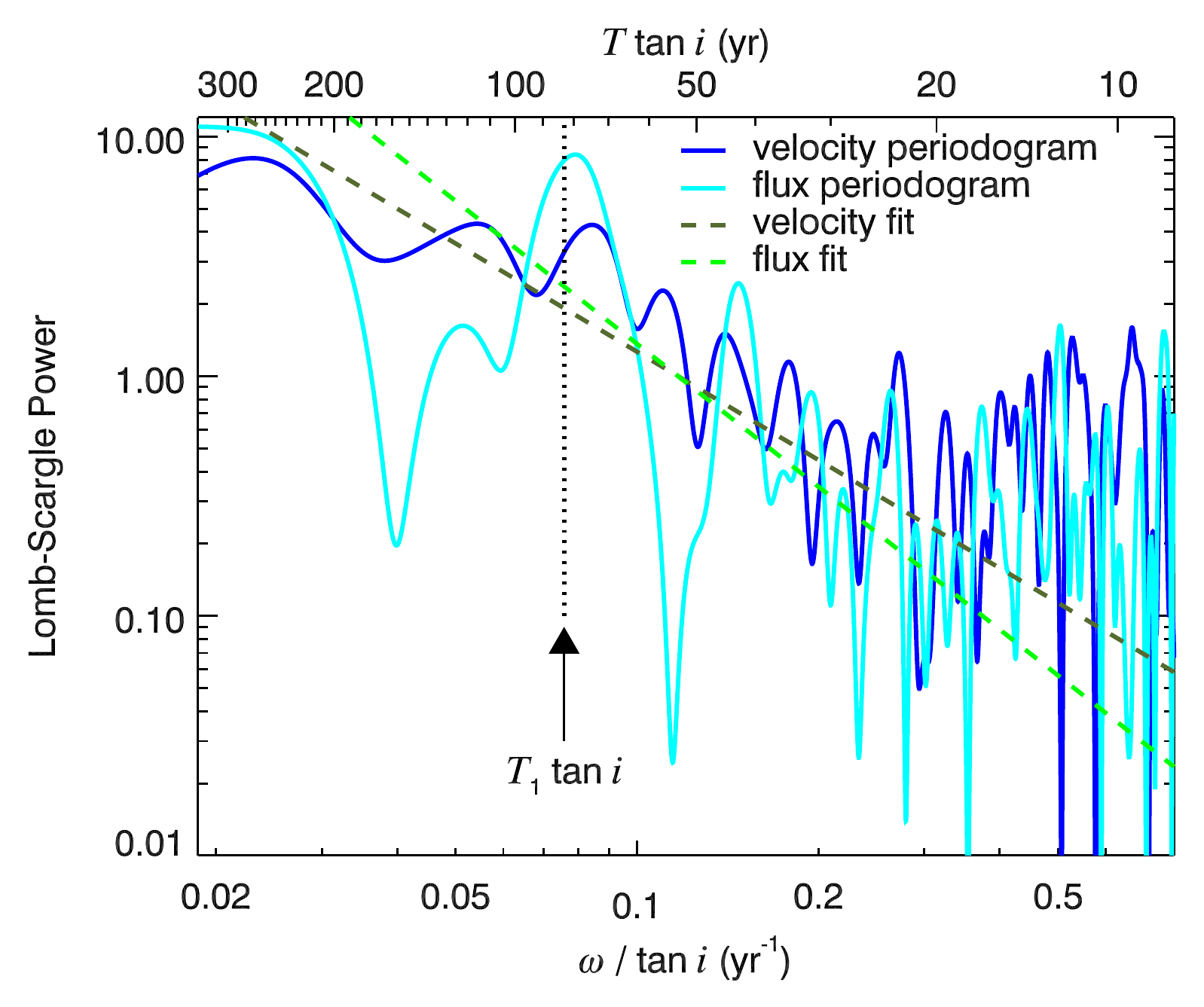}
   \caption{\textit{Top:} Timeseries ($\varv,t_{launch}$) and ($F,t_{launch}$) obtained by fitting the different Gaussian components in the emission profile (Fig.~\ref{fig:linean}) as described in Sect.~\ref{sec:analysis:measure} and applying Eq.~(\ref{eq:tlaunch}). 
\textit{Bottom:} Fourier transform (Lomb-Scargle periodogram) of the ($\varv,t_{launch}$) and ($F,t_{launch}$) coordinates in the blue lobe. The best-fit power-laws (dashed lines), minus the white noise constant $\gamma$, are overplotted. The dotted line indicates the largest period $T_1 \tan i$ in the \textit{periodic} model (Tab.~\ref{tab:modelparms});  the flux periodogram (and the velocity periodogram, albeit very weakly) also peaks at this value. 
   }
  \label{fig:fourier}
    \end{figure}

\subsubsection{Stochastic model}
\label{sec:analysis:stochastic}

In order to see whether a periodic signal is present in the time series data displayed in Fig.~\ref{fig:fourier} (top), we perform a Fourier analysis to the ($\varv \cos i,t_{\rm launch} \tan i$) coordinates and the ($F,t_{\rm launch} \tan i$) coordinates, where $F$ is the flux per gaussian component. Only the blue lobe data are fitted, as the signal is highest there. 

Fig.~\ref{fig:fourier} (bottom) shows the Lomb-Scargle periodogram \citep{Scargle1982} for the velocity and flux time series. There is no clear indication for a periodic signal in the velocity data, but increasing power at longer periods indicate the presence of a noise component. The flux data shows a peak at $\omega/\tan i \sim 0.08$~rad~yr$^{-1}$ ($T \tan i \sim 80$~yr) suggesting a (quasi-)periodic process on long timescales. Incidentally, this period coincides with the timescale found for the periodic model (see Sect.~\ref{sec:analysis:periodic}). We fit both periodograms, using a Maximum Likelihood approach appropriate for power spectra \citep[see e.g.][]{Vaughan2005}, with a power law plus a constant of the type
\begin{equation}
P(\nu) = \beta \nu^{-\alpha} + \gamma \; ,
\end{equation}
where $\alpha$ is the power law index, $\beta$ is a normalization term and $\gamma$ is a constant to account for the presence of white noise in the periodogram. The velocity periodogram is well fit with this model, with a power-law index of $\alpha = 1.50 \pm 0.04$. The flux periodogram is fitted by a power-law with index $\alpha=1.98 \pm 0.05$ but for the peak at $\sim 80$~yr.

In order to reconstruct the modulation in the outflow rate according to a noise process with the properties of periodograms for $v$ and $F$, we employed the method by \citet{Timmer1995} to simulate time series from periodogram data. We simulated 1000 time series $\varv(t)$ and $\dot{m}(t)$ from the fit to the velocity and flux periodograms, with a time resolution in the simulated time series of 10 years (which roughly corresponds to the spatial resolution of the detector). These time series served as input for the model described in Sect.~\ref{sec:analysis:model}.

\subsubsection{Periodic model}
\label{sec:analysis:periodic}

An alternative approach to reconstructing the modulation of the outflow rates $\varv(t)$, $\dot{m}(t)$ is by assuming them to be analytic, purely periodic functions:
\begin{align}
\varv(t) &= \varv_0+\varv_1 \sin 2\pi\left(\frac{t}{T_1}+\phi_1 \right);
\label{eq:vlaunch}\\
\dot{m}(t) &= \dot{m}_0 \left[1+ \sin 2\pi\left(\frac{t}{T_2} +\phi_2 \right) \right].
\label{eq:mlaunch}
\end{align}
Here, $T_{1}$ and $T_{2}$ are the periods of the oscillations in the velocity and mass outflow rate, respectively; $\phi_{1}$ and $\phi_{2}$ are the relative phases of the oscillations in the velocity and mass outflow rate, respectively. The parameters $\varv_{0}$ and $\varv_{1}$ are the mean and amplitude of the velocity variation; the parameter $\dot{m}_0$ is scalable and is normalized to fit the flux level in the data. As no clear symmetry exists between the knots in the red and blue lobe, a phase offset $\Delta \phi$ was introduced between the blue and red lobe outflow velocity. Summarizing, this amounts to seven free parameters for the periodic outflow rates: $\varv_0, \,\varv_1,\, T_1,\, T_2, \,\phi_1,\, \phi_2,$ and $\Delta\phi$.

A set of optimal values for these input parameters was obtained by fitting sinusoids to the data displayed in Fig.~\ref{fig:tfit} to the knot velocities (resulting in $\varv(t)$) and to the injection intervals (resulting in $\dot{m}(t)$). In order to fit the data, a phase offset between the red and blue lobes was introduced by setting $\Delta\phi = 0.65$. The model parameters listed in Tab.~\ref{tab:modelparms} thus obtained, result in a simulation that best represents the observed emission in position-velocity space. In the next section, a ballistic model is described with which we simulate the flow, with Eqs.~(\ref{eq:vlaunch}) and (\ref{eq:mlaunch}) as the input velocity and mass outflow rate. 

\begin{table}[!t]
\caption{\label{tab:modelparms}\normalsize{\textsc{Input parameters for periodic outflow rate, HH~1042.}}}
\begin{minipage}[c]{\columnwidth}
    \renewcommand{\footnoterule}{}
\centering
\begin{tabular}{llll}
\hline
\hline
Parameter & Value & Parameter & Value  \\
\hline
$\varv_0 \cos i$~\footnote{\scriptsize{The mean velocity $\varv_0 \cos i$ is fixed.}} & 170 km~s~$^{-1}$ & $\phi_1$ & $0.27$ (blue)  \\
$\varv_1 \cos i$ & 39.8 km~s~$^{-1}$ &  & $-0.38$ (red)\footnote{\scriptsize{The phase offset between the red and blue lobes is fixed at $\Delta \phi = 0.65$ in all simulations.}}\\
$T_1 \tan i$ & 83.0 yr & $\phi_2$ & $0.18$ (blue) \\
$T_2 \tan i$ & 12.0 yr & & $-0.47$ (red)\\
\hline
\vspace{-20pt}
\end{tabular}
\end{minipage}
\end{table}

\subsection{`Interaction physics': model for a ballistic flow}
\label{sec:analysis:model}

In this section we describe a one-dimensional model for the energy loss along a ballistic flow in a jet. This model is inspired by the approach of \citet{Raga1990}, who solved the inviscid Burgers equation for a variable flow in one dimension. In later work \citep{Raga2012} it was shown that the shock fronts (i.e. the knots) along such a flow merge by inelastic collisions. In our approach, we describe the flow in a Lagrangian picture, in terms of the collisions of discrete parcels of gas. As explained below, we parametrize between the limiting cases of a flow with inelastic collisions, and a free flow without collisions.

In our simulations, a sequence of discrete parcels of varying mass $m_i=\dot{m}\Delta t$ is ejected with a constant time interval $\Delta t$. The time interval is required to be small compared to the timescale on which the outflow rates vary. The distance traveled after $N$ discrete time steps $\Delta t$ by a parcel of gas launched at time $t_0$ is
\begin{equation}
x(t_N,t_0)=\sum_{i=0}^N \varv(t_i)\,\Delta t.
\label{eq:x}
\end{equation}


   \begin{figure*}[!ht]
   \centering
\includegraphics[width=\textwidth]{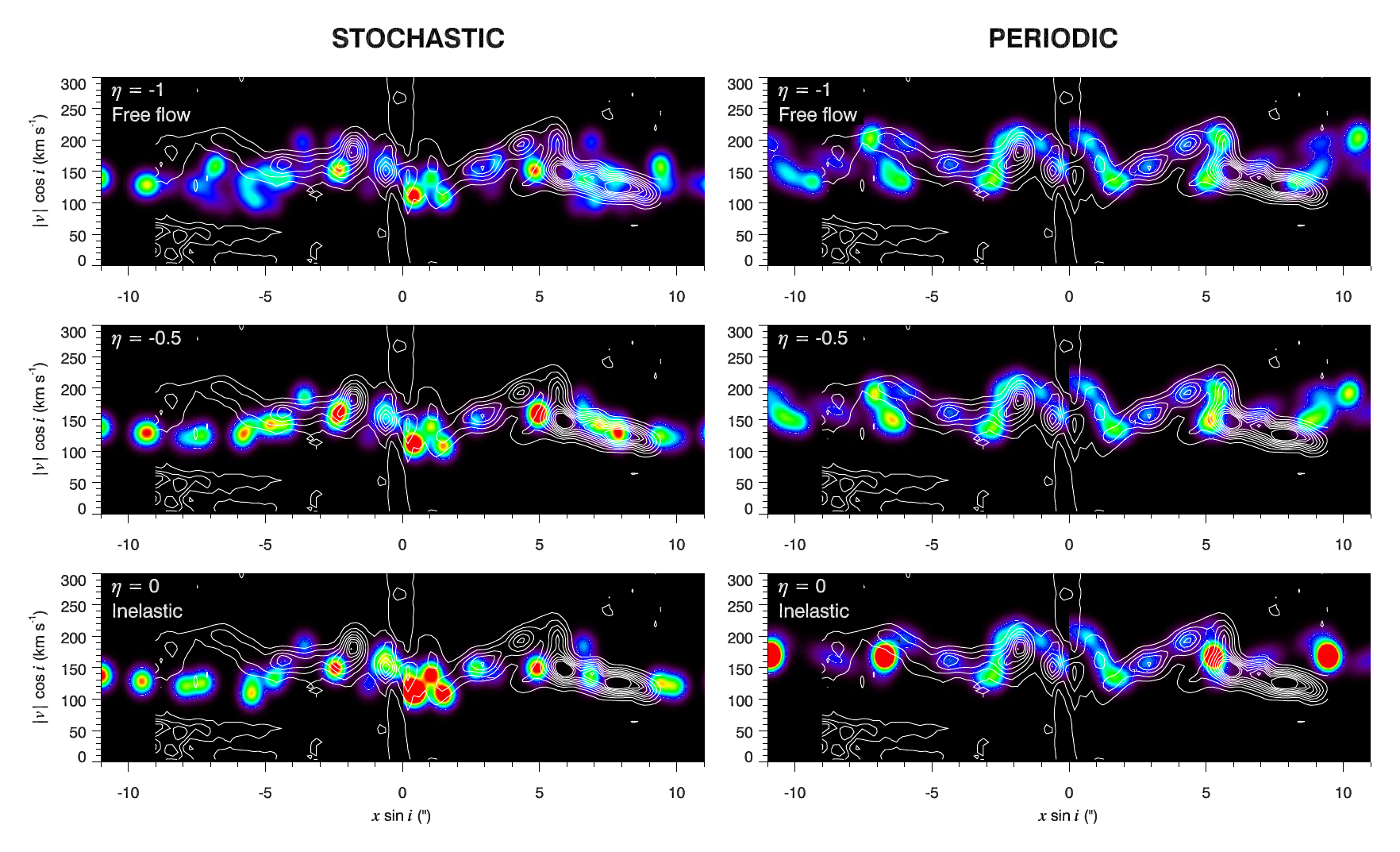}

   \caption{Results for the simulations run for HH~1042 for the stochastic (\textit{left}) and periodic (\textit{right}) models. The displayed stochastic result is selected from the 1000 generated stochastic models as it most adequately reproduces the observations. The periodic model parameters are listed in Tab.~\ref{tab:modelparms}. Colors indicate the modeled density field. White contours indicate the extinction-corrected observed [Fe\two]~1643~nm flux. The result is displayed for increasing values of the (global) energy loss parameter, $\eta = (-1, -0.5, 0)$ from top to bottom; $\eta=-1$ represents a free flow while $\eta=0$ represents a flow with fully inelastical collisions. Note the phase difference of $\Delta\phi=0.65$ between the red and blue lobe. }
\label{fig:simulationresults}
    \end{figure*}
    
\noindent When two parcels collide (i.e. a fast-moving parcel surpasses a slower moving parcel that was ejected earlier), momentum is conserved. It is possible that some kinetic energy is dissipated during a collision. Let us consider two parcels with masses $m_1$ and $m_2$, pre-shock velocities $\varv_1$ and $\varv_2$ and post-shock velocities $\varv'_1$ and $\varv'_2$. In order to calculate the energy loss, we perform a Galilean transformation to the center-of-momentum (COM) frame:
\begin{align}
\varv_1 &\rightarrow \varv_1 - V \equiv \tilde{\varv}_1\\
\varv_2 &\rightarrow \varv_2 - V \equiv \tilde{\varv}_2,
\end{align}
where tildes indicate quantities in the COM frame, and
\begin{equation}
V=\frac{m_1\varv_1+m_2\varv_2}{m_1+m_2}
\end{equation}
is the velocity of the COM. The momentum equation reads
\begin{equation}
\tilde{p} = \tilde{p}' = 0.
\end{equation}
We parametrize the energy loss \textit{in one collision} by the factor $\hat{\eta}^2$, defined such that
\begin{equation}
\tilde{E}'_{\rm kin} = \hat{\eta}^2 \tilde{E}_{\rm kin}, \label{eq:eta}
\end{equation}
where $0\le \hat{\eta}^2 \le 1$. Solving these equations for the velocities before and after the collision (assuming that no mass transfer occurs), and transforming back to the frame of reference of the observer, we have
\begin{align}
\varv'_1 = V \pm \hat{\eta}\,\frac{m_2(\varv_2-\varv_1)}{m_1+m_2} \label{eq:v1}\\
\varv'_2 = V \mp \hat{\eta}\,\frac{m_1(\varv_2-\varv_1)}{m_1+m_2}. \label{eq:v2}
\end{align}
We choose the solution with the plus sign in Eq.~(\ref{eq:v1}) and minus sign in Eq.~(\ref{eq:v2}), and require the energy loss parameter $\hat{\eta}$\footnote{$\hat{\eta}$ can be interpreted as the `coefficient of restitution' that is used in classical mechanics.} to take values between --1 and 0. A negative $\hat{\eta}$ can thus be interpreted as a `stickiness factor'. In our simulations, we have taken values between the two limiting cases $\hat{\eta}=-1$ (the two parcels pass each other without any interaction) and $\hat{\eta}=0$ (a fully inelastic collision occurs). The energy loss in one collision event is expressed in terms of the velocity difference and $\hat{\eta}$ as
\begin{equation}
\Delta E_{\rm kin} = \Delta \tilde{E}_{\rm kin}= \frac{1}{2}\mu'(1-\hat{\eta}^2)(\varv_1-\varv_2)^2,
\end{equation}
where $\mu' = m_1m_2/(m_1+m_2)$ is the reduced mass. 

Since a fraction $\hat{\eta}^2$ of the center-of-mass kinetic energy is dissipated in every collision, the total energy loss over a longer period scales exponentially with the number of collisions. In turn, this scales linearly with the number of particles generated per unit time in the simulation, which we wish to be an arbitrary parameter. We describe the energy loss in terms of a global parameter $\eta$:
\begin{equation}
\eta \equiv {\rm Sign}({\hat{\eta}})\,|\, \hat{\eta}\,|^{\Delta t/(0.5\, T_1)},
\end{equation}
with $0.5 \,T_1$ the characteristic timescale for variations in the flow velocity and $\Delta t$ the time step. In our simulations, $T_1$ is the longest period in the simulated outflow rate, Eq.~(\ref{eq:vlaunch}). In this definition, $\eta^2$ is a measure of the fraction of kinetic energy dissipated over one characteristic timescale. Using this global parameter, the results of the simulation is independent of the time step used. 

\subsection{Results of simulation}
\label{sec:analysis:simulationresults}

The simulations, which use the input and model described in the previous sections, consist of 500 time generations, running from $t \tan i=-320$~yr in the past up to the present day ($t \tan i=0$). To compare with observations, the absolute distances and velocites are converted to observable parameters ($x\sin i~[''], \varv\cos i$) by adopting $d$~=~0.7~kpc \citep{Liseau1992}. The fit parameters are thus degenerate with the inclination angle $i$. 

Because of the limitations of our simulations -- the ballistic approach, the inclusion of only one dimension, and the lack of a physical model that describes the density field, energy loss and emergent line emission -- it may only serve as an interpretative model. We thus limit ourselves to qualitative comparisons between data and simulations\footnote{We have run several tests to quantitatively compare the models to each other and to the data. However, no clear quantitative criteria can be established given the steep gradients present in the two-dimensional images.}.

In Fig.~\ref{fig:simulationresults} the results of the simulations are shown for both the stochastic and periodic models, for three different values of $\eta$, increasing from $\eta=-1$ up to $\eta=0$. Values of the parameter $\eta$ close to -1 lead to a larger spread in velocity, while $\eta$ close to zero leads to an increased density in the knots. Both the stochastic and the periodic outflow rates can produce a shocked density structure in the jet. 

The periodic model leads to a better representation of the observations than the stochastic model. The stochastic model displayed in Fig.~\ref{fig:simulationresults} is the one that out of the 1000 generated models most adequately represents the observations. While few of the stochastic models successfully reproduce the exact locations of the observed knots, the typical distance between them and their varying intensity is comparable to the observations.

The periodic rate produces a larger spread in velocity (e.g. the `saw-tooth' patterns in position-velocity space) which is generally not generated by the stochastic model. It reproduces the jet emission pattern in position-velocity space reasonably well. The locations of knots B, C, D and E in the blue lobe (and their substructure) are reproduced. The location of knot G (see Fig.~\ref{fig:sfr}) is also correctly predicted by the simulation. 

The validity of this interpretative model and its implications for constraints on the launching mechanism are further discussed in Secs.~\ref{sec:disc:model} and \ref{sec:disc:launchingmechanism}.

\section{Discussion}
\label{sec:disc}

In this section, we discuss the validity and context of our results. Sect.~\ref{sec:disc:mass} summarizes our findings on the mass and inclination angle of the two objects. In Sect.~\ref{sec:disc:physicalproperties} we compare the obtained physical characteristics of both sources with other YSOs across the mass spectrum. Finally, the method described in Sect.~\ref{sec:analysis} allows us to draw conclusions on the timescales, velocities, symmetry and collisions in the HH~1042 jet. These are summarized and discussed in Sect.~\ref{sec:disc:model}. Some scenarios describing the nature of the jet-launching mechanism within the constraints obtained from our analysis are put forward in Sect.~\ref{sec:disc:launchingmechanism}.

\subsection{Mass and inclination angle}
\label{sec:disc:mass}

Two important parameters are not well constrained: the mass of the central object, and the inclination of the disk-jet system. Knowing the mass of the central object is relevant because optical jets are rarely found around the more massive YSOs. The inclination is important because it remains as a free parameter in the mass loss estimates, as well as in the velocities, lengths and timescales derived in the kinematics analysis.

Lacking proper motion measurements, the best estimate of the inclination angle of the jets is represented by the inclination of the disk (assuming that the disk rotation axis is parallel to the jet). From fitting the CO-bandhead feature at 2.3~$\mu$m, the inclination angle of 08576nr292 is estimated at $i=27^{\circ+2}_{-14}$ by \citet{BikThi2004}, and at $i=17.8^{\circ+0.8}_{-0.4}$ by \citet{Wheelwright2010}. However, these results are dependent on the adopted stellar mass, because it sets the width of the modeled line profiles. The inclination angle in these models, keeping all other parameters fixed, then scales as $\sin i \propto M_*^{1/2}$. One should thus attempt to refine the estimate of $M_*$ to get a more accurate estimate of $i$.

The value $M_*= 6.1$~M$_\odot$ adopted in both studies mentioned above, based on the classification as an early B-type star by \citet{Bik2006}, is probably an overestimate. It does not take into account the intrinsic infrared excess that is evident from the SED-fitting by \citet{Ellerbroek2011}. From the extinction-corrected SED and the distance, and assuming the photosphere to be veiled by flat continuum emission from a disk, the stellar luminosity is expected to be between 10 and 70~$L_\odot$, equivalent to an A5V-B9V star with zero-age main sequence (ZAMS) radius. The lower limit is set by the luminosity integrated over the spectral range observed with X-shooter and \textit{Spitzer}/IRAC (Ellerbroek et al., in prep.); the upper limit is the brightest photosphere which could be `hidden' under the dereddened disk SED \citep{Ellerbroek2011}. The effective temperature probably is lower than the ZAMS value as PMS stars have bloated radii \citep[e.g.][]{Palla1993, Ochsendorf2011}. At its accretion rate ($\dot{M}_{\rm acc} \sim 10^{-6}$~M$_\odot$~yr$^{-1}$), the star is expected to arrive on the main sequence while still accreting \citep{Yorke2002}. However, the star is not expected to accrete more than a fraction of a solar mass over the remaining disk lifetime \citep{Hartmann1998}. Adopting the mass-luminosity relation for PMS stars \citep{Palla1993} and the considerations mentioned above, we tentatively estimate the final stellar mass of 08576nr292 at $M_* \sim 2-5$~M$_\odot$. Accounting for this large uncertainty in mass, we refine the estimate of the inclination angle to $i \sim 20^\circ-30^\circ$, which incidentally is consistent with the estimate of \citet{BikThi2004}. 

The spectrum of 08576nr480 does not contain signatures of its embedded stellar source other than the Ca\two~, O\one~and Br$\gamma$ lines, and the CO bandhead feature. Extinction prevents a classification of the optical SED, while at longer wavelengths ($3-10 \mu$m) the emission is probably contaminated by the surrounding cloud, preventing a classification of its infrared spectrum (Ellerbroek et al., in prep.). The jet appears to be quite inclined with respect to the line of sight due to the visibility of its two lobes and the bow-shock feature. For this reason, we have adopted $i=60^{\circ+15}_{-15}$ in Sect.~\ref{sec:mflux}. The CO bandhead of 08576nr480 is quite steep; a high inclination thus implies a low mass ($M \lesssim 1$) of the central object. The appearance of the jet is comparable to HH~1042 in terms of excitation conditions, and that more H$_2$ and H\one~lines are present in the nebular spectrum, suggesting the source is deeply embedded in the cloud and in an early evolutionary stage. This is consistent with an underestimated accretion rate (see Sect.~\ref{sec:disc:physicalproperties}).

From the $K$-band spectrum of HH~1043, it is apparent that a second continuum source is present at $\sim 0.8''$ (600 AU) westward of 08576nr480. Since the spectrum does not contain any stellar features, it cannot be said whether this is a companion or background star or scattered light of 08576nr480 off a clump in the jet.

\subsection{Physical properties, $\dot{M}_{\rm jet}$ and $\dot{M}_{\rm acc}$.}
\label{sec:disc:physicalproperties}

\begin{figure}[!t] 
   \centering
   \includegraphics[width=\columnwidth]{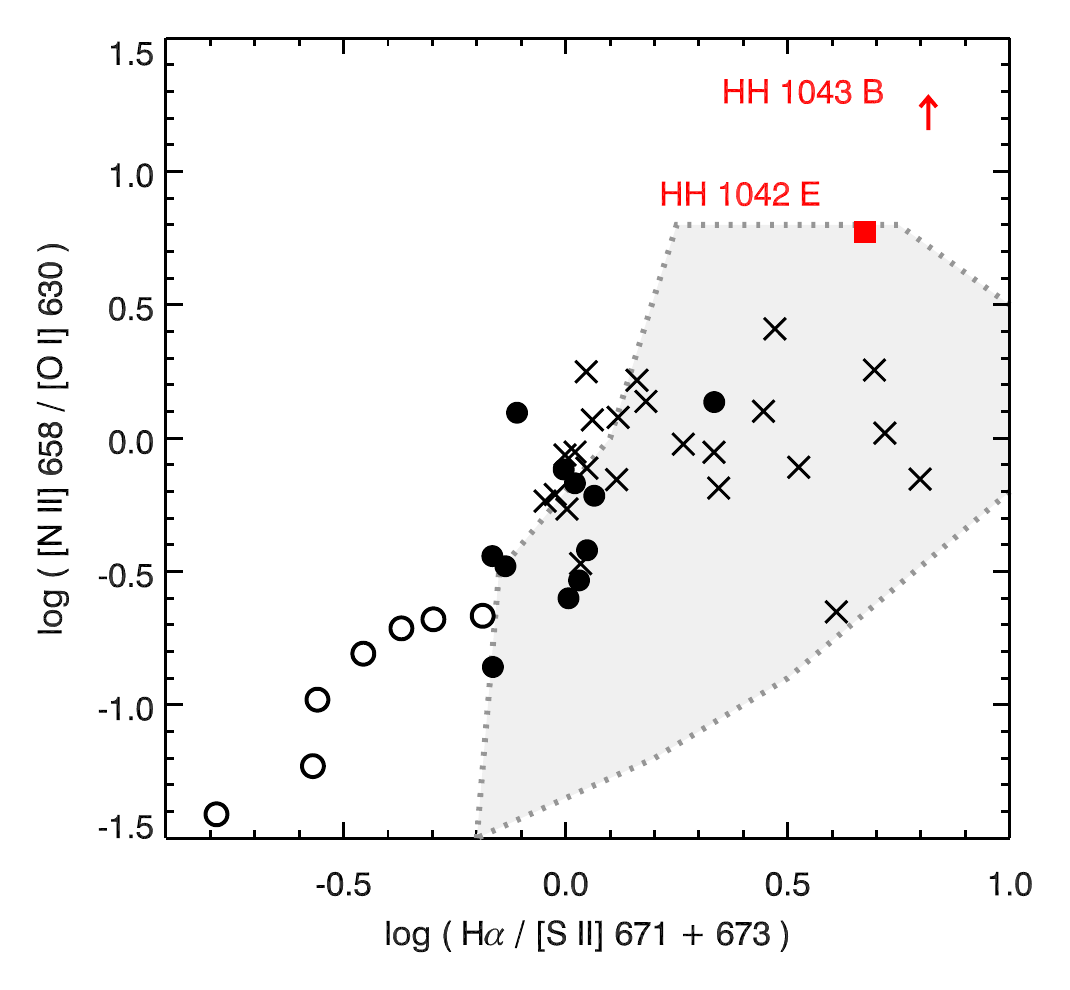} 
   \caption{Line ratios of HH sources with `low-' (open circles), `intermediate-' (filled circles) and `high-excitation' (crosses) spectra \citep[data and definitions adopted from][]{Raga1996}. Red symbols represent the brightest knots in HH~1042 (knot E) and HH~1043 (knot B). The shaded region denotates the ratios predicted by the plane-parallel shock and bow-shock models in \citet{Hartigan1987}. The high excitation in HH~1042 and HH~1043 can be explained by photoionization and/or -excitation by an external source of radiation.}
   \label{fig:ragaratios}
\end{figure}

The degree of excitation in both jet spectra is high compared to similar objects. Fig.~\ref{fig:ragaratios} depicts a diagram of two line ratios: H$\alpha$/[S\two] and [N\two]/[O\one], both known to be tracers of excitation \citep[see e.g.][]{Hartigan1994}. These line ratios in HH~1042 and HH~1043 are larger than those observed in `high-excitation' spectra \citep{Raga1996}, yet they are still within the range predicted by plane-parallel shock models by \citet{Hartigan1987}, given the large number of free parameters that these models depend on. However, the [S\three] lines in HH~1042 and HH~1043 are anomalously strong: two orders of magnitude stronger than in both models and observations of shocked jets. Apart from internal shocks, an external source of radiation may also contribute to these high ionization conditions. Candidates for this are either the relatively hot central source or the nearby O stars in RCW~36, or both.

   \begin{figure}[!ht]
\begin{minipage}{\columnwidth}

   \centering
\includegraphics[width=\columnwidth]{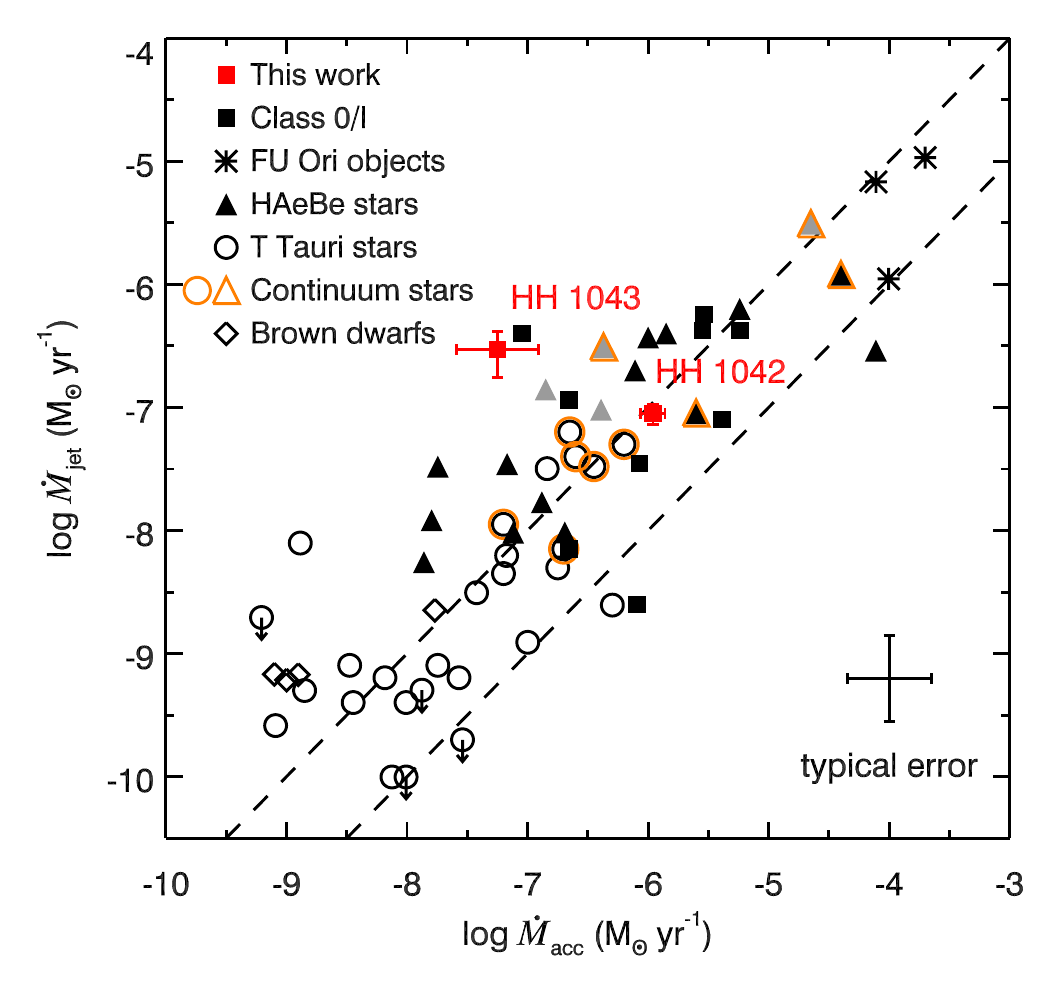}
   \caption[]{
   Observed mass outflow rate $\dot{M}_{\rm jet}$ versus accretion rate $\dot{M}_{\rm acc}$ of HH~1042 and HH~1043 (red squares) compared with different classes of YSOs associated with jets and outflows. The $\dot{M}_{\rm acc}$ value of some HAeBe stars (grey triangles) was determined from the H$\alpha$ luminosity in \citet{Hillenbrand1992}. The orange symbols indicate `continuum stars', objects classified as CTTS (orange circles) or HAeBe stars (orange triangles) with a high degree of veiling whose spectral type is very uncertain \citep[see][]{CalvetGullbring1998, Hernandez2004}; these objects are spectroscopically similar to 08576nr292, the driving source of HH~1042. The dashed lines indicate the $\dot{M}_{\rm jet}/\dot{M}_{\rm acc}=0.01$ and $0.1$ range, which is predicted by jet formation models. 
   
 \textit{References:} Class 0/I objects: \citet{Hartigan1994, Bacciotti1999, Podio2006, Podio2011, Podio2012, Antoniucci2008, Cabrit2009}; FU Orionis objects: \citet{Hartmann1995, Calvet1998}; HAeBe stars: \citet{Levreault1984, Boehm1994, Nisini1995, GarciaLopez2006, Wassell2006, Melnikov2008, Mendigutia2011, Donehew2011, Liu2011}; CTTS: \citet{Hartigan1995, Gullbring1998, Muzerolle1998b, Coffey2008, Cabrit2009, Melnikov2009}; Brown dwarfs: \citet{Whelan2007, Herczeg2008, Bacciotti2011}.
   }
  \label{fig:macc-mjet}
\end{minipage}
    \end{figure}

Fig.~\ref{fig:macc-mjet} displays the observed mass accretion and mass outflow rates of a sample of YSOs over a broad range in mass, evolutionary stage, and other properties. A correlation is seen between $\dot{M}_{\rm acc}$ and $\dot{M}_{\rm jet}$, albeit with a large scatter. Most sources are accretion-dominated ($\dot{M}_{\rm jet}/\dot{M}_{\rm acc} < 1$), while the accretion and outflow rates are highest in the more massive (HAeBe), young (Class 0/I) and extreme (FU Ori, continuum stars) objects. The mass outflow rates $\dot{M}_{\rm jet}$ are in some cases determined from spatially resolved jets, but mostly determined by unresolved on-source forbidden emission line diagnostics (cf. Sect.~\ref{sec:mflux}). The typical errors in the measurements of $\dot{M}_{\rm jet}$ and $\dot{M}_{\rm acc}$ are quite large, due to the uncertainty in the heating mechanisms and structure of the shocks, the structure of the accretion region, and the stellar parameters. These uncertainties can account for most of the scatter in Fig.~\ref{fig:macc-mjet}. Finally, the estimates of $\dot{M}_{\rm jet}$ and $\dot{M}_{\rm acc}$ differ greatly depending on the literature consulted, which may be due to the variable nature of the accretion/ejection process and/or the large uncertainties in the diagnostics used. 

The accretion and outflow rates of HH~1042 (08576nr292) are consistent with it being more massive than a CTTS and in a relatively early stage of evolution. The rates are higher than those observed in CTTS and are comparable with HAeBe-stars and the continuum stars, to which 08576nr292 bears many spectroscopic similarities. The accretion luminosity is of the same order as the stellar luminosity, which is consistent with models for intermediate-mass stars with high accretion rates \citep{Palla1993}. Our results imply a $\dot{M}_{\rm jet}/\dot{M}_{\rm acc}$ ratio of $\sim0.1$, consistent with values predicted by magneto-centrifugal forces \citep[see e.g.][]{Cabrit2009}. 

For HH~1043 (08576nr480), the estimated ratio $\dot{M}_{\rm jet}/\dot{M}_{\rm acc}$ exceeds unity. This indicates either an `outflow-dominated' system, or an underestimate of the accretion rate due to the accretion region being too embedded or obscured to be probed by the diagnostics used in Sect.~\ref{sec:macc} \citep[see e.g. the discussion in ][]{Bacciotti2011}. Based on the previous, 08576nr480 is likely a Class 0/I object. 

\subsection{Validity of the kinematics analysis and interpretative model}
\label{sec:disc:model}

The results of Sect.~\ref{sec:analysis} allow us to derive quantitative constraints on the launching mechanism of the HH~1042 jet (and to a lesser extent the HH~1043 jet), and also to comment qualitatively on its general outflow properties. 
First of all, the dynamical timescale (to be interpreted as the minimum jet lifetime, based on the mean velocity and spatial extent of the jets, taking into account the uncertainty in inclination) is 500~--~1000~yr for the blue lobe of HH~1042. The red lobe is visible up to 300~--~600~yr. For HH~1043 the dynamical timescales are 100~--~300~yr (blue lobe) and 100~--~400~yr (red lobe). The periods associated with the spatial intervals between the knots and the modulations of the outflow velocity are in the order of 10~--~100~yr in both jets. More specifically, the shape of HH~1042 in position-velocity space can be reconstructed reasonably well with periodic modulations of the outflow velocity $\varv(t)$ (with period $T_{1} \tan i = 83$~yr) and the mass outflow rate $\dot{m}(t)$ ($T_{2} \tan i = 12$~yr). The stochastic model produces similar knotted structure but a smaller velocity spread. The faint spike in the flux periodogram at $T \tan i \sim 80$~yr (Fig.~\ref{fig:fourier}) may reflect a periodic signal in the outflow rate. Incidentally, it coincides with the best-fit value $T_1 \tan i$ of the longest period in the periodic model. 

Instead of modulating $\dot{m}(t)$, one could also reproduce the knotted structure by superposing a second small amplitude modulation on $\varv(t)$, as done in e.g. \citet{Raga2012}. The resolution of our data does not allow us to distinguish between either a single-mode velocity and mass outflow modulation or a two-mode velocity modulation. In Fig.~\ref{fig:linean} (top) it can be seen that at the location of the knots, the velocity drops and the width of the individual components in the spectral profile increases. This suggests that they consist of different components scattered in velocity, which is not an uncommon observation in jets \citep{Hartigan2009}. While the detection of multiple shock fronts would put constraints on the launching mechanism, the spectral and spatial resolution of our observations are insufficient to test for their existence.

The mean velocity in HH~1042 is comparable in the blue and red lobe, $\varv_0 \cos i \sim 170$~km~s$^{-1}$. The amplitude of the modulations on this velocity is $\varv_1 \cos i \sim 40$ \kms. HH~1043 has mean velocities that are somewhat more asymmetric (80 and 60 \kms~in the blue and red lobes, respectively), but it could be that the system is moving with $10$~\kms~toward us with respect to the ambient cloud. Both the velocities and the timescales are similar to those derived for other jet sources \citep[e.g., HH~30, HH~34, HH~110, HH~111, HH~444, Hen~3-1475;][]{Raga2001, Raga2002a, Raga2002b, Raga2004, Raga2010, Raga2012, Velazquez2004, Esquivel2007}. Moreover, in most of these sources, like in the case of HH~1042, multiple modes with different timescales of order 10--1000~yr are found. Two timescales found in a single source typically have a ratio of order 10. These results may indicate that the mechanism responsible for jet launching is indeed intrinsically variable, and similar mechanisms causing variability may be at work in all HH objects across the stellar mass range. 

In both HH~1042 and HH~1043, the knots in the blue lobe do not have obvious counterparts in the red lobe. Consequently, there appears to be no correlation between ejection events in the blue and red lobes. A phase offset of $\Delta \phi = 0.65$ was adopted in the periodic models, but the parameter is not well constrained; a value of 0.5 cannot be excluded. The latter value would be a natural consequence of a precession movement of the jet (see the next subsection).

The simple ballistic model presented in this paper can qualitatively reproduce the structure observed in the jets. Shocks form along the jet wherever fast material catches up and collides with slower material. The local increase in density then results in an increased line emissivity \citep[e.g.][]{Bacciotti1999}. However, particularly in and beyond shock fronts (knot E), there is a mismatch in both velocity and flux level, indicating that a more sophisticated treatment of collision physics is needed. It is clear that some fraction of the kinetic energy is dissipated in collisions and either ionizes or excites the material, which heats and/or re-emits in different emission lines (\citealt[e.g.][]{Hartigan1994}; see Fig.~\ref{fig:pv3color}). The inclusion of a gas-dynamical model predicting line emissivities and fitting the data to more than one line species would likely improve these simulations, but is beyond the scope of this paper.

Models (both stochastic and periodic) with the energy loss parameter $\eta$ close to --1 barely have collisions and are therefore better suited to explain the velocity spread. However, it is evident from the variation in line emission (Fig.~\ref{fig:pv3color}) that ionization conditions vary along the jet, probably caused by shocks, which suggests that collisions do occur in the flow. In reality, the parameter $\eta$ may not have a single, global value as assumed in our models, but rather be dependent on local physical variables like density, temperature and degree of ionization.

\subsection{The nature of the jet-launching mechanism}
\label{sec:disc:launchingmechanism}

The outcome of the simulations in Sect.~\ref{sec:analysis} suggest that while the HH~1042 outflow rate may not be purely periodical, a quasi-periodic variation on the quoted typical timescales cannot be excluded. A straightforward explanation of the variability of velocities in the jet is an intrinsic variability of the accretion/ejection mechanism at the base of the jet. Variability of a quasi-periodic nature (i.e. with a characteristic, but not fixed timescale) can also be related to the piling up and subsequent release of material in the inner disk, or to the reconfiguration of the stellar magnetic field due to the rotation of the disk with respect to the star. 

In either of these scenarios, the origin of the phase shift is still unclear. It may be explained by a launching mechanism that does not produce knots in the red and blue lobes simultaneously. An alternative explanation for variability of the jet velocity is precession of the jet, which induces a half-period phase shift between the blue and red lobes. From the collimation seen in the SINFONI data (Fig.~\ref{fig:sfr}) the precession angle is constrained to be less than 10$^\circ$. The observed velocity mode $\varv_1$ could then be explained if the inclination of the system is $\sim 60^\circ$, corresponding to a $\lesssim 1$~M$_\odot$ star. Considering that the object is likely more massive, this scenario is not favored. A similar explanation for the phase lag is the movement of the jet on a binary orbit, which is not perfectly perpendicular to the jet axis. However, the values for the period and amplitude of this orbit render an unrealisticly high companion mass. Additional explanations for asymmetries in the launching mechanism are different ISM conditions or magnetic field configurations \citep{Matsakos2012}. However, these generally lead to a systematic asymmetry in the velocity of the blue and red lobes, while we observe a temporal asymmetry in the velocity modulations. 

\section{Summary}
\label{sec:conc}

We have presented a detailed study of the physical properties and kinematics of two newly discovered jets in RCW~36. These are interesting objects because of the visibility of both of their lobes, their being located in a massive star-forming region, the asymmetry in their kinematic structure, and the central source of HH~1042 being an intermediate-mass YSO. Below, the most important findings of this study are summarized.

\begin{itemize}

\item Both HH~1042 and HH~1043 display a shocked kinematic structure, with high ionization conditions in their terminal knots and at shock fronts along the jet (knot E in HH~1042). HH~1042 shows a `saw-tooth' pattern in position-velocity space, while the knots of HH~1043 are more widely separated spatially. The HH~1042 red lobe disappears into or beyond the molecular cloud, while its blue lobe terminates in an interaction region with the ISM (knot G). The HH~1043 blue lobe terminates in a clear bow-shock shaped structure (knot B).

\item The electron density in both jets is $n_{\rm e} \sim 10^3-10^4$~cm$^{-3}$. The electron temperature is $\sim 10^4$~K close to the source, then decreases further away from the source, similarly to what is found for low-mass YSOs. The ionization fraction is estimated to be low on-source ($<0.025$ in HH~1042~A), while it is very high in the outer knots ($\sim 0.7$ in HH~1042~E). 

\item The on-source extinction of 08576nr292 measured in the jet is several orders of magnitude lower than the value measured in the accretion column, suggesting the presence of circumstellar dust close to the star. The extinction in HH~1042 steeply increases in the red lobe, consistent with its disappearance into or behind a molecular cloud. 

\item Both jets have a mass outflow rate of order $\dot{M}_{\rm jet} \sim 10^{-7}$~M$_\odot$~yr$^{-1}$, comparable to what is found for HH objects from low-mass stars and more evolved intermediate-mass stars (HAeBe). The estimated accretion rate of HH~1042 ($\dot{M}_{\rm acc} = 1.10^{+0.29}_{-0.21}\times10^{-6}$~M$_\odot$~yr$^{-1}$) is high compared to CTTS accretion rates, consistent with a higher mass and/or the fact that its photosphere is heavily veiled by continuum emission from an optically thick accretion disk. The accretion rate of 08576nr480 (HH~1043) is more modest ($\dot{M}_{\rm acc} = 5.50^{+6.53}_{-2.99}\times10^{-8}$ M$_\odot$~yr$^{-1}$), yet it is likely underestimated as the source is embedded. 

\item For HH~1042, the combined optical and near-infrared diagnostics yield $\dot{M}_{\rm jet} / \dot{M}_{\rm acc} \sim 0.1$, which is consistent with values predicted by magneto-centrifugal forces \citep{Cabrit2009}, and with values observed in other, comparable sources. 

\item The kinematic structure of HH~1042 can be well simulated by an interpretative model with as inputs a stochastic or periodically variable outflow rate and a ballistic flow. The velocities (100 -- 200~km~s$^{-1}$) and derived timescales (10 -- 100 yr) are typical of those seen in other HH objects. 

\item A (quasi-)periodic signal is detected in the data at $T \tan i \sim 80$~yr (i.e. $T  \sim 100-200$~yr). The stochastic and periodic models both reproduce the knotted structure and apparent periodicity. The periodic model better reproduces the velocity spread in the data. 

\item The velocity spread is best reproduced by models with $\eta=-1$ (i.e. a small amount of kinetic energy dissipation). However, the amount of energy loss likely depends on local conditions in the flow.

\item The asymmetry between the positions and apparent launching times of the knots in the red and blue lobe of HH~1042 is well fitted by a phase shift of the outflow rate of 0.65 period, while a phase shift of 0.5 period cannot be excluded. Possible explanations for such a phase shift are an intrinsically asymmetric launching mechanism, or a precession movement of the jet. 

\item 08576nr292, the driving source of HH~1042, is an intermediate-mass YSO based on its jet and accretion properties. It resembles the class of `continuum stars': HAeBe-like stars with a photosphere veiled by extreme accretion activity. 

\end{itemize}

We have put forward various scenarios that might lead to the observed structure in the HH~1042 jet. Further high angular resolution and/or proper motion data may help to distinghuish between these. 

\acknowledgements{The anonymous referee is acknowledged for useful comments and suggestions that have improved the paper. The authors thank Daniele Malesani, Beate Stelzer and the ESO staff for obtaining some of the spectra. Francesca Bacciotti, Arjan Bik, Sylvie Cabrit, Carsten Dominik, Teresa Giannini, Michiel van der Klis, Lorenzo Maurri, Brunella Nisini, Rens Waters and Hugh Wheelwright are acknowledged for inspiring discussions.}


\appendix
\section{Additional materials}

\begin{table*}[!ht]
\begin{center}
\caption{\label{tab:flux1042blue}\normalsize{\textsc{Integrated fluxes per knot in HH~1042 (blue lobe).}}}

\hspace{0cm}\begin{tabular}{l l l l l l l l}
\hline
\hline
\multicolumn{2}{c}{}& \multicolumn{6}{c}{Flux (10$^{-16}$~erg~s$^{-1}$~cm$^{-2}$)}\\
Line & $\lambda$ (nm) & A & B & C & D & E &F  \\
\hline
$[$O\one$]$ &  630.0 & $     9.28 \pm     0.11 $  & $     1.43 \pm     0.09 $  & \dots  & \dots  & $     2.13 \pm     0.10 $  & \dots  \\
$[$O\one$]$ &  636.4 & $     2.21 \pm     0.10 $  & \dots  & \dots  & \dots  & \dots  & \dots  \\
$[$N\two$]$ &  654.8 & \dots  & \dots  & $     2.61 \pm     0.06 $  & $     1.41 \pm     0.04 $  & $     4.83 \pm     0.07 $  & $     3.37 \pm     0.07 $  \\
H$\alpha$ &  656.3 & $     16.9 \pm      0.1 $  & $     5.54 \pm     0.07 $  & $     19.8 \pm      0.1 $  & $     10.9 \pm      0.1 $  & $     36.7 \pm      0.1 $  & $     23.3 \pm      0.1 $  \\
$[$N\two$]$ &  658.3 & $     0.50 \pm     0.05 $  & $     2.07 \pm     0.04 $  & $     7.99 \pm     0.05 $  & $     4.45 \pm     0.04 $  & $     14.4 \pm      0.1 $  & $     7.26 \pm     0.07 $  \\
$[$S\two$]$ &  671.6 & $     0.52 \pm     0.05 $  & $     0.95 \pm     0.04 $  & $     1.40 \pm     0.04 $  & $     0.79 \pm     0.03 $  & $     2.92 \pm     0.05 $  & $     1.48 \pm     0.05 $  \\
$[$S\two$]$ &  673.1 & $     1.15 \pm     0.04 $  & $     1.44 \pm     0.04 $  & $     2.48 \pm     0.04 $  & $     1.43 \pm     0.03 $  & $     5.43 \pm     0.05 $  & $     2.88 \pm     0.06 $  \\
$[$Fe\two$]$ &  715.5 & $     4.35 \pm     0.04 $  & $     0.74 \pm     0.03 $  & $     0.76 \pm     0.03 $  & $     0.47 \pm     0.02 $  & $     1.56 \pm     0.03 $  & $     0.73 \pm     0.03 $  \\
$[$Ca\two$]$ &  729.1 & $     1.77 \pm     0.05 $  & $     0.31 \pm     0.04 $  & $     0.56 \pm     0.04 $  & $     0.35 \pm     0.03 $  & $     1.09 \pm     0.04 $  & $     0.46 \pm     0.04 $  \\
$[$O\two$]$ &  732.0 & \dots  & \dots  & $     0.79 \pm     0.04 $  & $     0.36 \pm     0.03 $  & $     1.33 \pm     0.04 $  & $     0.84 \pm     0.04 $  \\
$[$Ni\two$]$ &  737.8 & $     1.66 \pm     0.04 $  & $     0.52 \pm     0.03 $  & $     0.80 \pm     0.03 $  & $     0.50 \pm     0.02 $  & $     1.73 \pm     0.03 $  & $     0.74 \pm     0.03 $  \\
$[$Fe\two$]$ &  745.3 & $     2.07 \pm     0.04 $  & $     0.31 \pm     0.03 $  & \dots  & \dots  & $     0.58 \pm     0.02 $  & \dots  \\
O\one &  844.6 & $     5.69 \pm     0.05 $  & $     1.44 \pm     0.05 $  & \dots  & \dots  & $     1.60 \pm     0.02 $  & \dots  \\
Pa-14 &  859.8 & $     1.00 \pm     0.04 $  & \dots  & \dots  & \dots  & \dots  & \dots  \\
$[$Fe\two$]$ &  861.7 & $     10.3 $  & $     2.20 \pm     0.04 $  & $     2.28 \pm     0.03 $  & $     1.48 \pm     0.02 $  & $     5.75 \pm     0.03 $  & $     2.49 \pm     0.03 $  \\
Pa-12 &  875.0 & \dots  & \dots  & \dots  & \dots  & $     0.73 \pm     0.03 $  & \dots  \\
Pa-11 &  886.3 & $     2.98 \pm     0.06 $  & $     0.34 \pm     0.05 $  & \dots  & \dots  & $     1.01 \pm     0.03 $  & $     0.81 \pm     0.03 $  \\
$[$Fe\two$]$ &  889.2 & $     1.94 \pm     0.05 $  & $     0.69 \pm     0.05 $  & $     0.90 \pm     0.03 $  & $     0.52 \pm     0.02 $  & $     2.16 \pm     0.03 $  & $     1.21 \pm     0.03 $  \\
Pa-10 &  901.5 & $     5.46 \pm     0.07 $  & $     0.98 \pm     0.06 $  & $     0.97 \pm     0.03 $  & $     0.54 \pm     0.02 $  & $     2.00 \pm     0.03 $  & $     1.65 \pm     0.03 $  \\
$[$Fe\two$]$ &  905.2 & \dots  & \dots  & \dots  & \dots  & $     2.62 \pm     0.03 $  & \dots  \\
$[$S\three$]$ &  906.9 & $     1.28 \pm     0.07 $  & $     1.24 \pm     0.07 $  & $     6.75 \pm     0.04 $  & $     4.47 \pm     0.03 $  & $     15.3 $  & $     17.6 \pm      0.1 $  \\
Pa-9 &  922.9 & $     5.02 \pm     0.10 $  & $     1.10 \pm     0.09 $  & $     1.74 \pm     0.05 $  & $     0.93 \pm     0.03 $  & $     4.06 \pm     0.05 $  & $     3.22 \pm     0.05 $  \\
$[$S\three$]$ &  953.1 & \dots  & $     6.90 \pm     0.15 $  & $     40.8 \pm      0.1 $  & $     24.8 \pm      0.1 $  & $     88.9 \pm      0.2 $  & $     102. $  \\
Pa-8 &  954.6 & $     16.0 \pm      0.2 $  & \dots  & \dots  & \dots  & $     9.34 \pm     0.09 $  & \dots  \\
Pa$\delta$ & 1004.9 & $     12.1 \pm      0.2 $  & \dots  & \dots  & \dots  & $     9.36 \pm     0.12 $  & \dots  \\
He\one & 1083.0 & $     3.50 \pm     1.29 $  & \dots  & $     7.16 \pm     1.79 $  & $     9.72 \pm     1.34 $  & $     54.7 \pm      2.5 $  & $     92.0 \pm     14.6 $  \\
Pa$\gamma$ & 1093.8 & $     32.1 \pm      1.5 $  & $     3.17 \pm     1.20 $  & $     9.41 \pm     1.51 $  & $     5.80 \pm     1.12 $  & $     20.8 \pm      2.2 $  & $     23.7 \pm     18.4 $  \\
$[$Fe\two$]$ & 1256.7 & $     57.5 \pm      1.0 $  & $     31.7 \pm      0.8 $  & $     59.5 \pm      1.0 $  & $     45.5 \pm      0.8 $  & $     156. \pm       1. $  & $     83.6 \pm     11.8 $  \\
Pa$\beta$ & 1281.8 & $     81.0 \pm      1.5 $  & $     10.1 \pm      1.0 $  & $     27.5 \pm      1.3 $  & $     16.7 \pm      1.0 $  & $     67.1 \pm      1.5 $  & $     83.6 \pm     12.1 $  \\
$[$Fe\two$]$ & 1294.3 & $     14.7 \pm      1.6 $  & $     4.03 \pm     0.83 $  & $     9.00 \pm     0.97 $  & $     6.39 \pm     0.74 $  & $     22.0 \pm      1.1 $  & $     13.6 \pm     11.1 $  \\
$[$Fe\two$]$ & 1320.6 & $     47.9 \pm      1.4 $  & $     13.7 \pm      1.1 $  & $     24.9 \pm      1.3 $  & $     18.9 \pm      1.0 $  & $     64.7 \pm      1.5 $  & $     37.0 \pm     22.8 $  \\
$[$Fe\two$]$ & 1533.5 & $     12.1 \pm      0.8 $  & $     2.38 \pm     0.47 $  & $     7.58 \pm     0.54 $  & $     6.61 \pm     0.40 $  & $     21.4 \pm      0.6 $  & \dots  \\
$[$Fe\two$]$ & 1599.5 & $     18.8 \pm      0.8 $  & $     4.04 \pm     0.53 $  & $     7.74 \pm     0.70 $  & $     5.81 \pm     0.79 $  & $     22.1 \pm      0.8 $  & $     15.9 \pm      9.0 $  \\
Br-13 & 1610.9 & $     15.7 \pm      0.7 $  & \dots  & \dots  & \dots  & \dots  & \dots  \\
Br-12 & 1640.7 & $     18.0 \pm      0.6 $  & \dots  & \dots  & \dots  & \dots  & \dots  \\
$[$Fe\two$]$ & 1643.5 & $     59.7 \pm      0.5 $  & $     39.3 \pm      0.4 $  & $     75.0 \pm      0.5 $  & $     57.9 \pm      0.4 $  & $     205. \pm       1. $  & $     126. \pm       7. $  \\
$[$Fe\two$]$ & 1663.8 & $     5.90 \pm     0.64 $  & $     2.19 \pm     0.39 $  & $     4.56 \pm     0.43 $  & $     3.42 \pm     0.33 $  & $     12.8 \pm      0.5 $  & $     9.14 \pm     8.15 $  \\
Br-11 & 1680.7 & $     26.1 \pm      0.7 $  & \dots  & \dots  & \dots  & $     4.44 \pm     0.52 $  & \dots  \\
Br-10 & 1736.2 & $     52.3 \pm      0.7 $  & \dots  & \dots  & \dots  & \dots  & \dots  \\
Br-9 & 1817.4 & $     183. \pm      14. $  & $     17.1 \pm      7.9 $  & $     21.3 \pm      9.3 $  & $     7.92 \pm     6.87 $  & $     30.3 \pm     10.2 $  & \dots  \\
Br-8 & 1944.6 & $     594. \pm       3. $  & $     23.0 \pm      2.3 $  & $     20.2 \pm      2.8 $  & $     10.4 \pm      2.1 $  & $     42.7 \pm      3.1 $  & $     72.1 \pm     51.9 $  \\
He\one & 2058.1 & $     15.6 \pm      1.5 $  & \dots  & \dots  & \dots  & $     6.83 \pm     0.92 $  & \dots  \\
Br$\gamma$ & 2165.5 & $     103. \pm       1. $  & $     6.45 \pm     0.36 $  & $     13.2 \pm      0.4 $  & $     7.57 \pm     0.27 $  & $     31.2 \pm      0.4 $  & $     49.5 \pm      9.6 $  \\

\hline

\end{tabular}

\end{center}
\end{table*}%

\begin{table*}[!ht]
\begin{center}
\caption{\label{tab:flux1042red}\normalsize{\textsc{Integrated fluxes per knot in HH~1042 (red lobe).}}}

\hspace{0cm}\begin{tabular}{l l l l l l l }
\hline
\hline
\multicolumn{2}{c}{}& \multicolumn{5}{c}{Flux (10$^{-16}$~erg~s$^{-1}$~cm$^{-2}$)}\\
Line & $\lambda$ (nm) & A$'$ & B$'$ & C$'$ & D$'$ & E$'$  \\
\hline
$[$O\one$]$ &  630.0 & $     2.07 \pm     0.10 $  & \dots  & \dots  & \dots  & \dots  \\
$[$O\one$]$ &  636.4 & \dots  & \dots  & \dots  & \dots  & \dots  \\
$[$N\two$]$ &  654.8 & \dots  & \dots  & \dots  & \dots  & \dots  \\
H$\alpha$ &  656.3 & $     107. $  & $     2.26 \pm     0.05 $  & \dots  & \dots  & \dots  \\
$[$N\two$]$ &  658.3 & \dots  & \dots  & \dots  & \dots  & \dots  \\
$[$S\two$]$ &  671.6 & $     0.78 \pm     0.04 $  & $     1.05 \pm     0.05 $  & \dots  & \dots  & \dots  \\
$[$S\two$]$ &  673.1 & $     0.88 \pm     0.04 $  & $     1.43 \pm     0.04 $  & \dots  & \dots  & \dots  \\
$[$Fe\two$]$ &  715.5 & $     0.77 \pm     0.03 $  & \dots  & \dots  & \dots  & \dots  \\
$[$Ca\two$]$ &  729.1 & $     1.47 \pm     0.04 $  & \dots  & \dots  & \dots  & \dots  \\
$[$O\two$]$ &  732.0 & $     2.88 \pm     0.03 $  & \dots  & \dots  & \dots  & \dots  \\
$[$Ni\two$]$ &  737.8 & \dots  & \dots  & \dots  & \dots  & \dots  \\
$[$Fe\two$]$ &  745.3 & \dots  & \dots  & \dots  & \dots  & \dots  \\
O\one &  844.6 & $     12.1 $  & $     2.33 \pm     0.03 $  & \dots  & \dots  & \dots  \\
Pa-14 &  859.8 & $     3.18 \pm     0.02 $  & \dots  & \dots  & \dots  & \dots  \\
$[$Fe\two$]$ &  861.7 & $     2.83 \pm     0.02 $  & $     0.98 \pm     0.02 $  & \dots  & \dots  & \dots  \\
Pa-12 &  875.0 & $     6.07 \pm     0.03 $  & \dots  & \dots  & \dots  & \dots  \\
Pa-11 &  886.3 & $     11.8 $  & \dots  & \dots  & \dots  & \dots  \\
$[$Fe\two$]$ &  889.2 & \dots  & \dots  & \dots  & \dots  & \dots  \\
Pa-10 &  901.5 & $     11.0 $  & \dots  & \dots  & \dots  & \dots  \\
$[$Fe\two$]$ &  905.2 & \dots  & \dots  & \dots  & \dots  & \dots  \\
$[$S\three$]$ &  906.9 & \dots  & \dots  & \dots  & \dots  & \dots  \\
Pa-9 &  922.9 & $     14.6 $  & \dots  & \dots  & \dots  & \dots  \\
$[$S\three$]$ &  953.1 & \dots  & \dots  & \dots  & \dots  & \dots  \\
Pa-8 &  954.6 & $     51.2 \pm      0.1 $  & \dots  & \dots  & \dots  & \dots  \\
Pa$\delta$ & 1004.9 & $     39.5 \pm      0.1 $  & \dots  & \dots  & \dots  & \dots  \\
He\one & 1083.0 & $     62.1 \pm      1.5 $  & \dots  & \dots  & \dots  & \dots  \\
Pa$\gamma$ & 1093.8 & $     59.8 \pm      1.1 $  & \dots  & \dots  & \dots  & \dots  \\
$[$Fe\two$]$ & 1256.7 & $     19.2 \pm      1.0 $  & $     30.5 \pm      1.0 $  & $     8.82 \pm     0.87 $  & \dots  & \dots  \\
Pa$\beta$ & 1281.8 & $     166. \pm       1. $  & $     3.68 \pm     0.90 $  & \dots  & \dots  & \dots  \\
$[$Fe\two$]$ & 1294.3 & $     4.27 \pm     0.86 $  & $     3.33 \pm     0.62 $  & \dots  & \dots  & \dots  \\
$[$Fe\two$]$ & 1320.6 & $     19.4 \pm      1.0 $  & $     10.7 \pm      0.7 $  & $     3.04 \pm     0.57 $  & \dots  & \dots  \\
$[$Fe\two$]$ & 1533.5 & $     24.1 \pm      0.7 $  & $     6.19 \pm     0.32 $  & $     1.91 \pm     0.25 $  & \dots  & \dots  \\
$[$Fe\two$]$ & 1599.5 & $     6.16 \pm     0.59 $  & $     3.37 \pm     0.35 $  & \dots  & \dots  & \dots  \\
Br-13 & 1610.9 & $     3.36 \pm     0.49 $  & \dots  & \dots  & \dots  & \dots  \\
Br-12 & 1640.7 & $     9.79 \pm     0.49 $  & \dots  & \dots  & \dots  & \dots  \\
$[$Fe\two$]$ & 1643.5 & $     21.9 \pm      0.5 $  & $     40.7 \pm      0.3 $  & $     13.3 \pm      0.2 $  & $     8.71 \pm     0.28 $  & \dots  \\
$[$Fe\two$]$ & 1663.8 & $     4.10 \pm     0.53 $  & \dots  & \dots  & \dots  & \dots  \\
Br-11 & 1680.7 & $     9.83 \pm     0.53 $  & \dots  & \dots  & \dots  & \dots  \\
Br-10 & 1736.2 & $     21.7 \pm      0.5 $  & \dots  & \dots  & \dots  & \dots  \\
Br-9 & 1817.4 & $    2797. \pm     268. $  & $     268. \pm     129. $  & \dots  & \dots  & \dots  \\
Br-8 & 1944.6 & $     268. \pm       2. $  & $     5.52 \pm     0.92 $  & \dots  & \dots  & \dots  \\
He\one & 2058.1 & $     13.9 \pm      0.9 $  & \dots  & \dots  & \dots  & \dots  \\
Br$\gamma$ & 2165.5 & $     99.8 \pm      0.6 $  & \dots  & \dots  & \dots  & \dots  \\

\hline

\end{tabular}

\end{center}
\end{table*}%

%

\begin{table*}[!ht]
\begin{center}
\caption{\label{tab:flux1043}\normalsize{\textsc{Integrated fluxes per knot in HH~1043 (blue and red lobes).}}}

\hspace{0cm}\begin{tabular}{l l l l l l l l}
\hline
\hline
\multicolumn{2}{c}{}& \multicolumn{6}{c}{Flux (10$^{-16}$~erg~s$^{-1}$~cm$^{-2}$)}\\
\multicolumn{2}{c}{}& \multicolumn{2}{c}{\textit{Blue lobe}}& \multicolumn{4}{c}{\textit{Red lobe}}\\
Line & $\lambda$ (nm) & A & B & A$'$ & B$'$ & C$'$ & D$'$  \\
\hline
$[$O\one$]$ &  630.0 & \dots  & \dots  & \dots  & \dots  & \dots  & \dots  \\
$[$O\one$]$ &  636.4 & \dots  & \dots  & \dots  & \dots  & \dots  & \dots  \\
$[$N\two$]$ &  654.8 & \dots  & \dots  & \dots  & \dots  & \dots  & \dots  \\
H$\alpha$ &  656.3 & $     30.4 \pm      0.1 $  & $     40.3 \pm      0.1 $  & $     25.2 \pm      0.1 $  & $     21.9 \pm      0.1 $  & $     6.54 \pm     0.06 $  & $     3.83 \pm     0.06 $  \\
$[$N\two$]$ &  658.3 & $     6.97 \pm     0.07 $  & $     14.3 \pm      0.1 $  & $     9.20 \pm     0.07 $  & $     7.09 \pm     0.07 $  & $     1.60 \pm     0.04 $  & $     1.12 \pm     0.04 $  \\
$[$S\two$]$ &  671.6 & \dots  & $     2.31 \pm     0.05 $  & $     1.10 \pm     0.05 $  & $     0.79 \pm     0.05 $  & \dots  & \dots  \\
$[$S\two$]$ &  673.1 & $     1.50 \pm     0.05 $  & $     4.47 \pm     0.05 $  & $     2.05 \pm     0.05 $  & $     1.38 \pm     0.05 $  & $     0.29 \pm     0.03 $  & \dots  \\
$[$Fe\two$]$ &  715.5 & \dots  & \dots  & \dots  & \dots  & \dots  & \dots  \\
$[$Ca\two$]$ &  729.1 & \dots  & \dots  & \dots  & \dots  & \dots  & \dots  \\
$[$O\two$]$ &  732.0 & $     1.75 \pm     0.05 $  & $     3.56 \pm     0.05 $  & $     2.31 \pm     0.03 $  & $     1.27 \pm     0.03 $  & $     0.28 \pm     0.02 $  & \dots  \\
$[$Ni\two$]$ &  737.8 & \dots  & \dots  & \dots  & \dots  & \dots  & \dots  \\
$[$Fe\two$]$ &  745.3 & \dots  & \dots  & \dots  & \dots  & \dots  & \dots  \\
O\one &  844.6 & \dots  & \dots  & \dots  & \dots  & \dots  & \dots  \\
Pa-14 &  859.8 & \dots  & \dots  & \dots  & \dots  & \dots  & \dots  \\
$[$Fe\two$]$ &  861.7 & $     1.88 \pm     0.03 $  & $     2.09 \pm     0.02 $  & $     1.93 \pm     0.03 $  & $     1.18 \pm     0.03 $  & \dots  & \dots  \\
Pa-12 &  875.0 & \dots  & \dots  & \dots  & \dots  & \dots  & \dots  \\
Pa-11 &  886.3 & \dots  & \dots  & \dots  & \dots  & \dots  & \dots  \\
$[$Fe\two$]$ &  889.2 & \dots  & \dots  & \dots  & \dots  & \dots  & \dots  \\
Pa-10 &  901.5 & \dots  & $     2.07 \pm     0.03 $  & $     1.76 \pm     0.03 $  & $     1.56 \pm     0.03 $  & \dots  & \dots  \\
$[$Fe\two$]$ &  905.2 & \dots  & \dots  & \dots  & \dots  & \dots  & \dots  \\
$[$S\three$]$ &  906.9 & $     19.3 \pm      0.1 $  & $     19.6 \pm      0.1 $  & $     33.6 \pm      0.1 $  & $     40.7 \pm      0.1 $  & $     14.2 \pm      0.0 $  & $     8.20 \pm     0.04 $  \\
Pa-9 &  922.9 & $     3.77 \pm     0.07 $  & $     4.15 \pm     0.06 $  & $     3.04 \pm     0.06 $  & $     2.64 \pm     0.05 $  & \dots  & \dots  \\
$[$S\three$]$ &  953.1 & $     138. $  & $     169. $  & $     109. $  & $     101. $  & $     37.1 \pm      0.1 $  & $     24.9 \pm      0.1 $  \\
Pa-8 &  954.6 & \dots  & \dots  & \dots  & $     8.26 \pm     0.09 $  & \dots  & \dots  \\
Pa$\delta$ & 1004.9 & \dots  & \dots  & $     11.7 \pm      0.1 $  & $     11.0 \pm      0.1 $  & $     3.44 \pm     0.07 $  & \dots  \\
He\one & 1083.0 & $     198. \pm       2. $  & $     212. \pm       2. $  & $     172. \pm       2. $  & $     155. \pm       2. $  & $     66.9 \pm      1.0 $  & $     54.5 \pm      1.1 $  \\
Pa$\gamma$ & 1093.8 & $     28.0 \pm      1.6 $  & $     27.3 \pm      1.4 $  & $     22.0 \pm      1.0 $  & $     21.2 \pm      0.9 $  & $     5.18 \pm     0.57 $  & $     4.90 \pm     0.61 $  \\
$[$Fe\two$]$ & 1256.7 & $     59.1 \pm      1.0 $  & $     68.9 \pm      0.9 $  & $     64.7 \pm      1.4 $  & $     55.5 \pm      1.2 $  & $     6.53 \pm     0.51 $  & $     18.2 \pm      0.6 $  \\
Pa$\beta$ & 1281.8 & $     94.5 \pm      1.5 $  & $     84.1 \pm      1.4 $  & $     92.6 \pm      1.3 $  & $     88.0 \pm      1.0 $  & $     21.8 \pm      0.6 $  & $     20.2 \pm      0.9 $  \\
$[$Fe\two$]$ & 1294.3 & $     9.28 \pm     0.98 $  & $     10.2 \pm      0.9 $  & $     12.8 \pm      0.7 $  & $     9.11 \pm     0.61 $  & \dots  & $     3.30 \pm     0.55 $  \\
$[$Fe\two$]$ & 1320.6 & $     24.0 \pm      1.0 $  & $     24.0 \pm      1.0 $  & $     20.9 \pm      0.9 $  & $     18.5 \pm      1.0 $  & $     2.12 \pm     0.45 $  & $     6.25 \pm     0.65 $  \\
$[$Fe\two$]$ & 1533.5 & $     15.1 \pm      0.6 $  & $     16.8 \pm      0.6 $  & $     24.2 \pm      0.4 $  & $     19.6 \pm      0.3 $  & $     2.86 \pm     0.19 $  & $     7.07 \pm     0.39 $  \\
$[$Fe\two$]$ & 1599.5 & $     14.3 \pm      0.9 $  & $     12.4 \pm      0.7 $  & $     16.1 \pm      0.5 $  & $     11.9 \pm      0.3 $  & $     1.26 \pm     0.21 $  & $     4.75 \pm     0.34 $  \\
Br-13 & 1610.9 & $     5.07 \pm     0.47 $  & \dots  & $     3.85 \pm     0.53 $  & $     4.09 \pm     0.29 $  & \dots  & $     1.10 \pm     0.27 $  \\
Br-12 & 1640.7 & $     5.80 \pm     0.38 $  & $     4.75 \pm     0.34 $  & $     4.52 \pm     0.32 $  & $     4.91 \pm     0.27 $  & $     1.12 \pm     0.17 $  & $     1.12 \pm     0.21 $  \\
$[$Fe\two$]$ & 1643.5 & $     95.4 \pm      0.9 $  & $     103. $  & $     103. $  & $     90.1 \pm      0.7 $  & $     10.1 \pm      0.2 $  & $     33.7 \pm      0.3 $  \\
$[$Fe\two$]$ & 1663.8 & $     7.13 \pm     0.42 $  & $     7.13 \pm     0.37 $  & $     9.25 \pm     0.41 $  & $     7.18 \pm     0.28 $  & \dots  & $     2.95 \pm     0.31 $  \\
Br-11 & 1680.7 & $     13.2 \pm      0.5 $  & $     12.9 \pm      0.4 $  & $     4.00 \pm     0.38 $  & $     5.20 \pm     0.28 $  & $     1.27 \pm     0.17 $  & $     1.62 \pm     0.36 $  \\
Br-10 & 1736.2 & $     14.4 \pm      0.5 $  & $     9.88 \pm     0.68 $  & $     12.0 \pm      0.3 $  & $     11.8 \pm      0.2 $  & $     2.91 \pm     0.14 $  & $     3.26 \pm     0.27 $  \\
Br-9 & 1817.4 & $     46.9 \pm     14.1 $  & $     35.9 \pm      9.3 $  & $     244. \pm     222. $  & \dots  & \dots  & \dots  \\
Br-8 & 1944.6 & $     337. \pm      24. $  & $     410. \pm      22. $  & $     42.9 \pm      1.1 $  & $     52.3 \pm      0.9 $  & $     11.6 \pm      0.6 $  & $     13.5 \pm      1.0 $  \\
He\one & 2058.1 & $     47.3 \pm      0.9 $  & $     25.6 \pm      1.0 $  & $     31.8 \pm      0.9 $  & $     33.7 \pm      0.5 $  & $     11.5 \pm      0.3 $  & $     11.1 \pm      0.6 $  \\
Br$\gamma$ & 2165.5 & $     80.9 \pm      0.6 $  & $     63.2 \pm      0.5 $  & $     74.6 \pm      0.4 $  & $     78.2 \pm      0.4 $  & $     16.5 \pm      0.3 $  & $     20.5 \pm      0.4 $  \\

\hline

\end{tabular}

\end{center}
\end{table*}

\end{document}